\documentclass{cernyrep}

\usepackage{varwidth}

\usepackage{xcolor}

\newcommand\query[1]{%

\colorbox{yellow}{\begin{varwidth}{\dimexpr\linewidth-2\fboxsep}\textbf{[AQ:~{#1}]}\end{varwidth}}}
\usepackage{amsbsy}

\begin{document}

\title{Beam-driven, Plasma-based Particle Accelerators}
\author{P. Muggli}
\institute{Max Planck Institute for Physics, München, Germany}
\maketitle

\begin{abstract}
We briefly give some of the characteristics of the beam-driven, plasma-based particle accelerator known as the plasma wakefield accelerator (PWFA).
We also mention some of the major results that have been obtained since the birth of the concept. %
We focus on high-energy particle beams where possible.
\end{abstract}

\query{Please supply keywords}
%


\section{Introduction}
In plasma-based particle accelerators (PBPAs), particles are accelerated by the wakefields sustained by a periodic plasma density perturbation.
It was first realized that wakefields can be driven in plasmas by intense laser pulses~\cite{tajima79}.
In this case, it is the ponderomotive force associated with the gradient of the laser pulse intensity and not the oscillating electric field itself that acts to displace the plasma electrons and drive the wakefields.
Such a PBPA
is called a laser wakefield accelerator (LWFA).

Soon after this, it was realized that an intense, relativistic charged-particle bunch can also drive the wakefields~\cite{chen85}.
In this case, the (unipolar) space charge field of the bunch acts on the plasma electrons.
A PBPA driven by a charged-particle bunch is called a plasma wakefield accelerator (PWFA).
This is a particular type of a collinear wakefield accelerator.

The PWFA is one of the advanced accelerator schemes studied as a high-gradient alternative to today's RF technology.

It is, of course, hopeless to pretend to summarize PWFA research in a few pages.
However, we attempt to touch on some of the most relevant points, to give an introduction to the field.
The many details missing here can be found in the references provided and in other articles.
It is left to the reader to do the detailed work, which is the essence of the learning process.

The text is organized as follows.
First, we outline a few characteristics of charged particles and of charged-particle bunches relevant for the PWFA.
Then we briefly describe how wakefields are driven in the plasma. We next introduce the concept of the transformer ratio. After that, we summarize two-dimensional PWFA linear theory and the relevance of the longitudinal and transverse dimensions of the bunch to the driving process.
We also introduce the concepts of the wave-breaking field and beam loading.
We then introduce the non-linear regime of the PWFA, including beam focusing, propagation and acceleration, both for electron and positron bunches.
The possibility of using a hollow plasma channel for positron bunches is also briefly addressed.
We mention the intermediate, quasi-linear or weakly non-linear regime of the PWFA, as well as the self-modulation instability.
We end with a few remarks.

\section{Charged particles, charged-particle bunches and the PWFA}

Charged-particle bunches have a number of characteristics that make them particularly suitable for driving wakefields for PBPAs. %
\begin{itemize}
\item The electric field of a relativistic charge or bunch is essentially transverse. %
This can be seen by using the Lorentz transform for the purely radial electric field of a single particle in its rest frame: $E_{r0}=\frac{q}{4\pi\epsilon_0}\frac{1}{r^2}$. %
In the laboratory frame where the particle has relativistic factor $\gamma$, the fields are $E_{||}=E_{r0}$, $E_{\perp}=\gamma E_{r0}$, $B_{||}=0$, $B_{\perp}=(v_\mathrm{b}/c^2)E_{r0}$. %
Therefore, $E_{\perp}=\gamma E_{||}\gg E_{||}$ for $\gamma\gg1$. %
\item A charged-particle bunch experiences a space charge force that makes it diverge (transversely) and lengthen (longitudinally). %
However, when the particles are relativistic, their relative distance does not change significantly over distances of interest (a few metres for PBPAs). %
For particles with energies $\gamma$ and $\gamma+\Delta\gamma$, with $\Delta\gamma\ll\gamma$, the dephasing $\Delta L$ over a propagation distance $L$ is $\Delta L/L\cong(1/\gamma^2)(\Delta\gamma/\gamma)$. %
This dephasing is usually small compared with the length of interest, the plasma wavelength $\lambda_{\mathrm{pe}}=2\pi c/\omega_{\mathrm{pe}}$ over plasma lengths utilized in PBPAs. %
There is, therefore, no lengthening of the relativistic bunch under the conditions of interest here. %
In the transverse plane, the bunch is subject to the full Lorenz force, $F_{\perp}=q_\mathrm{b}(E_{\perp}+v_\mathrm{b}\times B_{\perp})$. %
Here, $v_\mathrm{b}=\left(1-1/\gamma^2\right)^{1/2}c$ is the velocity of the bunch particles; in cylindrical coordinates, $v_\mathrm{b}$ is along the $z$-axis and $E_{\perp}=E_r$, $B_{\perp}=B_{\theta}$. %
To evaluate the fields, we need to assume a cylindrical  infinitely long \emph{bunch} or beam, with uniform density $n_\mathrm{b}$. %
In this case, the radial electric field within the bunch radius is simply given by Gauss' law: $E_r=\frac{1}{2}\frac{q_\mathrm{b}n_\mathrm{b}}{\epsilon_0}r$ . %
The magnetic field is given by Faraday's law, $B_{\theta}=\frac{1}{2}(\mu_0q_\mathrm{b}n_\mathrm{b})r$, with $q_\mathrm{b}=\pm e$ the charge of a bunch particle. %
The total force therefore reads, with $c^2=1/\mu_0\epsilon_0$,
\begin{equation}
F_{\perp}=q_\mathrm{b}(E_{r}+v_\mathrm{b}\times B_{\theta})=q_b\frac{1}{2}\frac{q_\mathrm{b}n_\mathrm{b}}{\epsilon_0}\left(1-\frac{v_\mathrm{b}^2}{c^2}\right)r=\frac{1}{\gamma^2}q_\mathrm{b}E_r\ . %
\label{eqn:Fperp}
\end{equation}
That is, the total transverse force is reduced to $1/\gamma^2$ multiplied by the pure space charge force $F_{\mathrm{sc}}=q_\mathrm{b}E_r$, \ie as soon as the particles become relativistic ($\gamma\gg1$), the transverse dynamics are dominated by the emittance and the external focusing forces. %
\item A beam with geometric emittance $\epsilon_\mathrm{g}$ has a beta-function at its waist, defined as $\beta_0=\sigma_0^2/\epsilon_\mathrm{g}$, where its transverse r.m.s.\ size is $\sigma_0$. %
This parameter depends on the beam emittance, which can, in principle, be made very small. %
The beta-function of a charged-particle beam is the equivalent of the Rayleigh length of a photon beam focused to a transverse size $w_0$: $Z_\mathrm{R}=\pi\frac{w_0^2}{\lambda_0}$, where $\lambda_0$ is the wavelength of the laser pulse. %
In both cases, the transverse size of the beam increases by a factor of $\sqrt 2$ over $\beta_0$ or $Z_\mathrm{R}$.
Since the vast majority of today's laser pulses that are short and intense enough to drive a PBPA have $\lambda_0\cong800$\Unm, the Rayleigh length is determined only by the focal size, $w_0$. %
For example, choosing a typical value $\sigma_0=w_0=$10\Uum, we obtain $\beta_0=0.8 (0.08)$\Um\ (for a typical Stanford Linear Accelerator Center (SLAC) beam geometric emittance, $\epsilon_\mathrm{g}=1.26\times10^{-9}(1.26\times10^{-10})$\Um\Urad) in the $x$-($y$-)plane and $Z_\mathrm{R}=400\Uum$. %
The value of $\beta_0$ and $Z_\mathrm{R}$ determines the distance over which the beam remains small and can thus drive wakefields (in the absence of external forces or a guiding mechanism). %
A future linear collider beam is expected to have an $\epsilon_\mathrm{g}$ of the order of $(1-8)\times10^{-11}$\Um \Urad\ from $200$ and $500\UGeV$~\cite{bib:ILC}.


\item The space charge field of the bunch is (at least) partially cancelled by the plasma (see P. Gibon's lecture). %
Its current is also cancelled by the plasma return current. %
Even an initially radially uniform plasma, therefore, focuses the bunch and counters its natural divergence, resulting from its non-zero incoming emittance. %
In addition to the beam having a relatively long beta-function, plasma focusing can maintain a small beam transverse size and the wakefield driving and acceleration over long distances. %
\item The velocity of the particle bunch in the plasma is independent of the plasma density. %
For a bunch with particle energy $E_0$ and relativistic factor $\gamma=E_0/m_\mathrm{e}c^2-1\propto E_0/m_\mathrm{e}c^2$ it is simply given by the particle's velocity: $v_\mathrm{b}=\left(1-1/\gamma^2\right)^{1/2}c$. %
The laser pulse velocity is given by the group velocity of the light in the plasma and around frequency $\omega_0$: $v_\mathrm{g}=\left(1-\omega_{\mathrm{pe}}^2/\omega_0^2\right)^{1/2}c\le c$. %
Here $\omega_{\mathrm{pe}}=\left(n_{\mathrm{e}0}e^2/\epsilon_0m_\mathrm{e}\right)^{1/2}$ is the plasma electron (angular) frequency in a plasma of (electron) density $n_{\mathrm{e}0}$. %
\end{itemize}

In the next section, we describe the principles of the PWFA. %

\section{The PWFA}

\begin{figure}[ht]
\centering
\includegraphics[width=12cm]{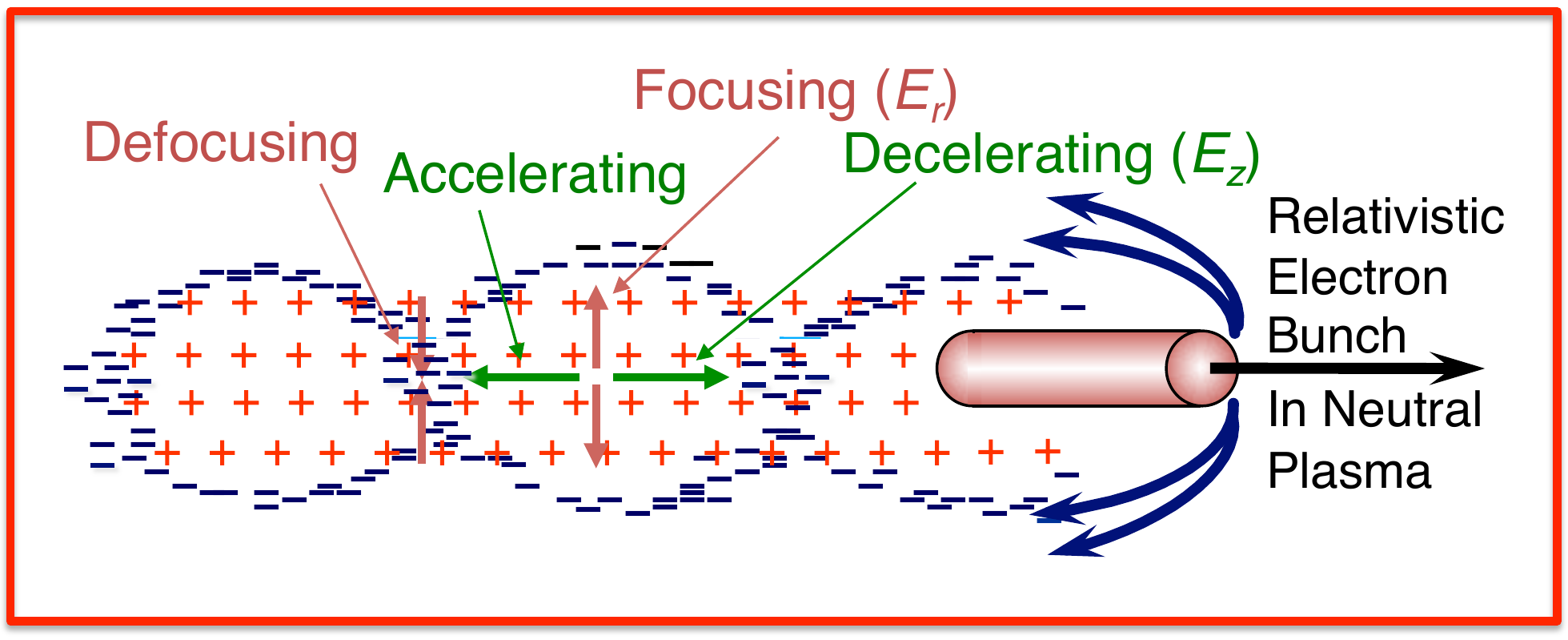}
\caption{Schematic of the PWFA principle. %
The electron bunch displaces the plasma electrons and forms regions of the plasma that are globally positively and negatively charged. %
This electron plasma density perturbation sustains the wakefields. %
The main electric field components of the wakefields are shown and can be schematically determined from the charge distribution. %
This is the case of a bunch with density larger than the plasma density ($n_\mathrm{b}\gg n_{\mathrm{e}0}$, non-linear PWFA regime) and a pure ion column is formed. %
The arrows showing the backward motion of the plasma electrons suggest this as seen in a window moving with the drive bunch. }
\query{Some of these figures appear to have been taken wholesale from other published work. Please check that you have obtained any necessary permission to reproduce figures. Please ensure that  this is clearly marked in the appropriate figure caption,  e.g. `Reproduced with permission from  (name of copyright holder), etc., and that the source material is clearly and explicitly referenced.            }
\label{fig:PWFAscematic1}

\end{figure}

In the PWFA, the mostly transverse space charge field of the relativistic charged-particle bunch travelling in a neutral plasma displaces the plasma electrons (Fig.~\ref{fig:PWFAscematic1}). %
The positively or negatively charged bunch driving the wakefields is called the drive bunch. %
The plasma ions experience the same force as the electrons but, because of their larger mass or inertia, respond on a much longer time scale, of the order of the inverse of the ion plasma frequency, $\omega_{\mathrm{pi}}=\left(n_{\mathrm{i}0}Z^2e^2/\epsilon_0M_i\right)^{1/2}$. %
In an initially neutral plasma, the ion density is $n_{\mathrm{i}0}=n_{\mathrm{i}0}$. %
Here, $Z$ is the number of ionized electrons per atom ($Z=1$ for the protons of a hydrogen plasma or singly ionized atom with more electrons) and $M_\mathrm{i}$ is the ion mass. %
Ions are thus usually considered as immobile over the typical $\omega_{\mathrm{pe}}^{-1}$ time scale of a single wakefield period since $\omega_{\mathrm{pe}}\gg\omega_{\mathrm{pi}}$.
When the drive bunch density is much larger than the plasma density, the ions can move over the same time scale as the electrons and ion motion must be considered~\cite{ionsmotion,ionsmotion2,ionsmotion3}. %

In the case of a negatively charged drive bunch (\eg an electron bunch), once displaced, the plasma electrons feel the restoring force of the plasma ions, are attracted back to the axis, overshoot and oscillate. %
This oscillation of the plasma electrons with period $\cong2\pi/\omega_{\mathrm{pe}}$ and phased by the drive bunch moving at approximately the speed of light is the plasma wake. %
The direction of the fields can be determined form the charge distribution in the second accelerator structure or \emph{bubble} (see Fig.~\ref{fig:PWFAscematic1}). %
On the axis there is an alternation of positively and negatively charged regions. %
In the front of the structure, the longitudinal $E_z$ field is in the forward direction (for the bunch moving to the right), corresponding to a decelerating field or force. %
The drive bunch particles lose energy to the plasma in expelling the plasma electrons (for the case of the negatively charged drive bunch). %
In the back of the structure, the $E_z$ field direction is reversed, allowing for the possibility for particles in the back of the bunch, or in a trailing witness bunch, to gain energy from the wakefields. %
In the middle of the structure, the transverse field points outwards, corresponding to a focusing force for this drive bunch or for a witness bunch of the same charge sign. %
Between the structures, regions of strongly compressed plasma electron density correspond to defocusing regions. %
\par The wakefield can be driven (energy loss) and sampled (energy gain) by the electrons of a single bunch, approximately one plasma wavelength long, as suggested by Fig.~\ref{fig:PWFAscematic1} and as demonstrated next. %
However, this leads to a large final energy spread, since all phases of the wakefields are sampled. %
\par
\begin{figure}[ht]
\begin{center}
\includegraphics[width=12cm]{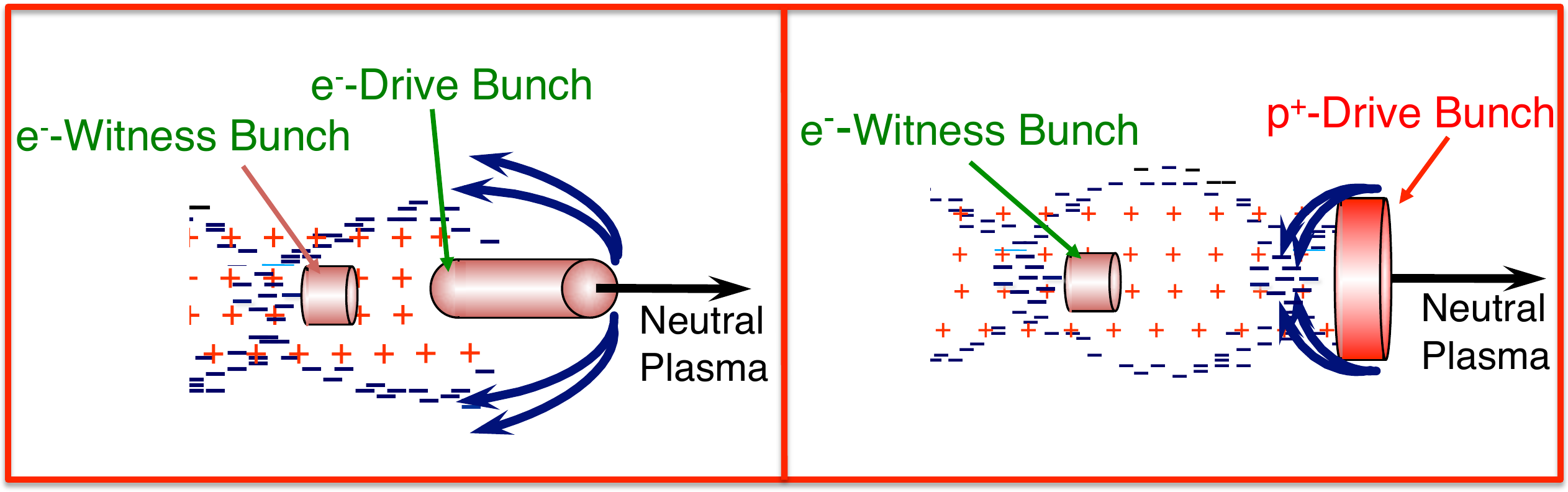}
\caption{The left panel shows the wakefields driven by a negatively charged drive bunch. %
The witness bunch that is accelerated is also negatively charged. %
The right panel shows a schematic of the wakefields driven by a positively charged drive bunch. %
The witness bunch is negatively charged. %
}
\label{fig:PWFAscematic2}
\end{center}
\end{figure}

 Wakefields can be driven by a shorter bunch and sampled by another short, trailing bunch, called a witness bunch. %
This is shown in Fig.~\ref{fig:PWFAscematic2}. %
This can lead to a narrow final energy spectrum for the witness bunch, as demonstrated in Ref.~\cite{muggli09} and recently in Ref.~\cite{bib:litos14}. %
\par The drive bunch can also be positively charged, as suggested in Fig.~\ref{fig:PWFAscematic2}. %
In this case, the plasma electrons are first attracted towards the beam axis, but then sustain the same kind of wakefield, with just a phase shift with respect to the negatively charged drive bunch. %

Driving wakefields with a charged particle bunch was first demonstrated in Ref.~\cite{bib:rosenzweig88}, interestingly, with a drive-witness electron bunch train. %
Driving wakefields with a positively charged particle bunch was first demonstrated with positrons in Ref.~\cite{blue}. %

The witness bunch can also have a positive charge (\eg positrons). %
In this case, it must be placed in the corresponding accelerating and focusing phase (region) of the wakefields. %

\section{Transformer ratio}

The transformer ratio is an important concept for the PWFA. %
It is the ratio of the accelerating field amplitude behind the drive bunch(es) to the decelerating field amplitude \emph{within} the drive bunch(es). %
It can be defined from general wakefield characteristics (\ie also in RF systems). %

Consider the wakefield amplitude per unit charge $W(\xi)$ at a position $\xi$ at or behind a particle. %
Following Ref.~\cite{ruth84}, the rate of energy change (loss in this case) by an infinitely short bunch with $N_1$ charges $e$ of energy $E_1$ (per particle) located at  $\xi=0$ is
\begin{equation}
\frac{\mathrm{d}(N_1E_1)}{\mathrm{d}z}=-\left(N_1e\right)N_1eW(0)=-N_1^2e^2W(0)\ .
\label{eqn:rate1}
\end{equation}
Similarly, the rate of energy change of the second bunch with charge $N_2e$ and energy $E_2$ located at  $\xi=\xi_2$ is
\begin{equation}
\frac{\mathrm{d}N_2E_2)}{\mathrm{d}z}=-N_2^2e^2W(0)+\left(N_2e\right)N_1eW(\xi_2)=-N_2^2e^2W(0)+N_1N_2e^2W(\xi_2)\ .
\label{eqn:rate2}
\end{equation}
Note that $N_1$ and $N_2$ remain constant along the plasma and that the result is independent of the sign of the charge $e$; it depends on the sign of $W(\xi_2)$, \ie on the position (or phase) of the second bunch in the wakefield of the first one. %
The sum of the energy change by the two bunches must be smaller or equal to zero, thus
\begin{equation}
\frac{\mathrm{d}(N_1E_1)}{\mathrm{d}z}+\frac{\mathrm{d}(N_2E_2)}{\mathrm{d}z}=-N_1^2e^2W(0)-N_2^2e^2W(0)+N_1N_2e^2W(\xi_2)\le0\ ,
\label{eqn:totalrate}
\end{equation}
or
\begin{equation}
\left(N_1^2+N_2^2\right)W(0)+N_1N_2W(\xi_2)\ge0\ .
\label{eqn:totalrate2}
\end{equation}
Since this relation has to be true for all values of $N_1$ and $N_2$, we obtain
\begin{equation}
2W(0)\le -W(\xi_2)\ .
\label{eqn:inequal}
\end{equation}
The accelerating gradient $G$ is, from Eqs.~(\ref{eqn:rate2}) and~(\ref{eqn:inequal}),
\begin{equation}
G=\frac{\mathrm{d}E_2}{\mathrm{d}z}=-N_2e^2W(0)+N_1e^2W(\xi_2)\le-N_2e^2W(0)+2N_1e^2W(0)=e^2W(0)\left(2N_1-N_2\right)\ .
\label{eqn:gain}
\end{equation}
The distance over which the first bunch loses all its energy is
\begin{equation}
N_1E_1=N_1^2e^2W(0)L\ ,
\label{eqn:grad1}
\end{equation}
or
\begin{equation}
L=\frac{E_1}{N_1e^2W(0)}\ .
\label{eqn:Lplasma}
\end{equation}
Thus, the change in energy of the trailing particles is
\begin{equation}
\Delta E_2=GL\le\frac{e^2W(0)}{N_1e^2W(0)}E_1\ ,
\label{eqn:grad2}
\end{equation}
and
\begin{equation}
\Delta E_2\le E_1\left(2-\frac{N_2}{N_1}\right)\ .
\label{eqn:E2}
\end{equation}
This expression is a maximum for $N_2=0$, in which case $\Delta E_2\le2E_1$, showing that the energy gain per trailing particle is at most equal to twice the incoming energy of a drive particle. %
Since both drive and witness bunches travel the same distance in the plasma, the transformer ratio can be defined from the longitudinal wakefield amplitude: $R=E_{+}/E_{-}$. %
Here, $E_{-}$ is the maximum decelerating field \emph{within} the drive bunch and $E_{+}$ is the maximum accelerating field \emph{behind} the drive bunch, where the particles of the witness bunch can be placed. %
Note that global energy conservation implies that $N_2\Delta E_2\le N_1E_1$, or, more generally, $N_2\Delta E_2\le N_1\Delta E_1$ over a distance smaller than or equal to the full energy depletion length defined in Eq.~(\ref{eqn:Lplasma}). %
It is also important to note that these expressions relate the energy change (loss/gain) of the different particles. %
This means that, for example, a bunch of $20\UGeV$ particles could lose (almost, they must remain relativistic) all their energy (\ie lose $20\UGeV$ per particle) and transfer $40\UGeV$ to each particle of a witness bunch of $100\UGeV$. %
This is what would happen in a staged accelerator, where each plasma section (stage) would be driven by a $20\UGeV$ bunch, while a witness bunch would keep gaining (up to) $40\UGeV$ per stage. %

This concept is similar to that for an electrical transformer, $U_2I_2\le U_1I_1$ (index 1 for primary and 2 for secondary; this expression is written for power rather than energy), where the number of particles per bunch divided by the bunch length is the current. %
The PWFA is a transformer in which the energy of a high-charge, low-energy bunch is extracted and added to that of a lower-charge witness bunch, through the plasma wakefields. %

It is interesting to notice that the transformer ratio concept is valid for all collinear accelerators and that the expressions were derived from very simple and general assumptions of energy conservation and linear superposition of the wakefields. %
In real systems, the bunches can evolve in the wakefields and other considerations that come into play (beam loading, \etc) that can reduce the transformer ratio reached. %

Transformer ratios much larger than two can be obtained by tailoring the bunch longitudinal density or current profile, or by using a sequence of bunches~\cite{laziev}. 
The transformer ratio has been measured in PWFA experiments~\cite{bib:blumenfeld10}. %

\section{PWFA linear regime}

The linear regime is interesting because in this regime many fundamental aspects of the PWFA can be calculated directly. %
It is, a priori, not interesting for a collider because (by definition) it does not use the full potential of the plasma for sustaining large fields. %
These large fields are reached when the plasma perturbation is also large, \ie in the non-linear regime of the PWFA. %
In addition, in this regime the wakefields have continuous longitudinal and radial variations and the beam and plasma characteristics are directly proportional to each other (by definition). %
Therefore, the characteristics of the system generally evolve along the propagation distance, and are not necessarily suitable for a long-lasting acceleration process that aims to produce a high-quality bunch with a small final energy spread and emittance. %

\subsection{Linear theory}
The linear theory of the PWFA can be derived using a cold, non-relativistic fluid model for the plasma. %
A clear derivation can be found in Ref.~\cite{chen87}. %
This model uses Newton's equation for a fluid element, the continuity equation and Poisson's equation. %
The equations are linearized (see examples in Ref.~\cite{ffchen} for the linearization process) and a wave equation for the plasma electron density perturbation $n_1\ll n_{\mathrm{e}0}$ can be derived. %
It is driven by the bunch density $n_\mathrm{b}$ and reads
\begin{equation}
\frac{\partial^2n_1}{\partial \xi^2}+k_{\mathrm{pe}}^2n_1=\frac{q_\mathrm{b}}{e}k_{\mathrm{pe}}^2n_\mathrm{b}\ ,
\label{eqn:wave}
\end{equation}
where $n_\mathrm{b}\ll n_{\mathrm{e}0}$ is assumed. %
These equations are written in the coordinate system of the bunch, \ie $\xi=z-v_\mathrm{b}t$, which is often written as $\xi=z-ct$ for relativistic bunches. %
Equations are written as a function of space coordinates ($z$ and $\xi$) rather than the time coordinate, a more natural choice for a linear accelerator. %
In these coordinates, $\frac{\mathrm{d}}{\mathrm{d}z}=\frac{\mathrm{d}}{\mathrm{d}\xi}$ and $\frac{\mathrm{d}}{\mathrm{d}t}=-v_b\frac{\mathrm{d}}{\mathrm{d}\xi}\cong-c\frac{\mathrm{d}}{\mathrm{d}\xi}$. %
Note the term related to the bunch charge (\eg $q_\mathrm{b}/e=\mp1$ for an electron or positron bunch), which is the source term driving the (harmonic) oscillator. %
Note also that a similar equation can be obtained for the case of a laser pulse driving the plasma density perturbation with the ponderomotive force as the source term. %
This equation can be solved in 1D for a delta function bunch or charge to obtain the Green's function for the wakefield generation ($n_1$ perturbation) and the longitudinal wakefields using Poisson's equation. %
This longitudinal component has a $\cos\left(k_{\mathrm{pe}}\xi\right)$ dependency. %
The radial dependency can be obtained from 2D theory for a given transverse density profile (Gaussian, parabolic, \etc) and involves Bessel functions~\cite{chenfoc} and a $\sin\left(k_{\mathrm{pe}}\xi\right)$ dependency. %
For a bunch with longitudinal and radial Gaussian profiles, the wakefields read
\begin{equation}\label{eq:MultiBunches_long1}
W_z (\xi, r)=\frac{e}{\epsilon_{0}}\int_{-\infty}^{\xi} n_{\mathrm{b}_{\parallel}}(\xi')\cos\left[k_{\mathrm{pe}}(\xi-\xi')\right]\mathrm{d}\xi'\cdot R(r)\ ,
\end{equation}
\begin{equation}\label{eq:MultiBunches_tran1}
W_{\perp} (\xi, r)=\frac{e}{\epsilon_{0}k_{\mathrm{pe}}}\int_{-\infty}^{\xi} n_{\mathrm{b}_{\parallel}}(\xi')\sin\left[k_{\mathrm{pe}}(\xi-\xi')\right]\mathrm{d}\xi'\cdot \frac{\mathrm{d}R(r)}{\mathrm{d}r}\ ,
\end{equation}
where $R(r)$ is the transverse dependency given by
\begin{equation}\label{eq:MultiBunches_Rr}
\begin{array}{rcl}
R(r)={k_{\mathrm{pe}}}^2\int_0^rr'\,\mathrm{d}r'n_{b\perp}(r')I_0(k_{\mathrm{pe}}r')K_0(k_{\mathrm{pe}}r)+{k_{\mathrm{pe}}}^2\int_r^\infty r'\mathrm{d}r'n_{b\perp}(r')I_0(k_{\mathrm{pe}}r)K_0(k_{\mathrm{pe}}r')\ ,
 \end {array}
\end{equation}
where $I_0$ and $K_0$ are the zeroth-order modified Bessel functions of the first and second kind, respectively. %
Note that the $W$ notation is typical of wakefields in structures. %
It is important to understand the meaning of these equations. %

On the one hand, in linear theory, $W_z$ is associated only with an $E_z$ component, since $v_\mathrm{b}$ is along the $z$-axis and the Lorentz force has no magnetic contribution along $z$ (to first, linear order). %
On the other hand, $W_{\perp}$ has two components and is often written as $E_r-v_\mathrm{b}B_{\theta}$ in cylindrical coordinates. %
The terms have a $\xi$ dependency that is a pure 1D, longitudinal dependency. %
This dependency is obtained from the Green's function for wakefield excitation. %
The $R(r)$ and $\mathrm{d}R/\mathrm{d}r$ terms express the deviation from 1D theory obtained in 2D. %
The longitudinal (accelerating) wakefield is a maximum on the axis with $R(r=0)$ (see~\cite{katsou}). %
For $k_{\mathrm{pe}}\sigma_r\gg1$, \ie $\sigma_r\gg c/\omega_{\mathrm{pe}}$, $R(r=0)\rightarrow 1$, i.e., one recovers the 1D result, as expected. %
For $k_{\mathrm{pe}}\sigma_r\ll1$, i.e., $\sigma_r\ll c/\omega_{\mathrm{pe}}$, $R(r=0)\rightarrow k_{\mathrm{pe}}^2\sigma_r^2\left(0.0597-\ln(k_{\mathrm{pe}}\sigma_r)\right)$. %
This means that when the plasma density is decreased for a fixed $\sigma_r$ (or $\sigma_r$ for a fixed $n_{\mathrm{e}0}$, and in both cases a fixed bunch charge) the radial component contributes to a decrease in $W_z$, even if $n_\mathrm{b}/n_{\mathrm{e}0}$ increases. %
This is because 1D theory applies to an infinite sheet bunch (in the perpendicular direction). %
In 2D, the plasma responds with a natural transverse extent of (a few) $c/\omega_{\mathrm{pe}}$. %
Therefore, when $k_{\mathrm{pe}}\sigma_r\ll1$, the charge contained in the (few) $c/\omega_{\mathrm{pe}}$ is less than in the 1D case and so is $W_z$ (for a given charge density).

A few important remarks about these equations: %
\begin{itemize}
\item They are symmetrical with respect to the bunch charge sign, which means that, in the linear regime, the wakefields are the same for both charge signs, to within a phase factor of $\pi/2$.
%

\item The wakefields are necessarily initially decelerating within the drive bunch because it displaces the plasma electrons and has to lose energy in the process. %
The wakefields are necessarily initially focusing within the drive bunch because the plasma electrons move to shield the bunch fields, \ie to neutralize the bunch charge and decrease the space charge fields. %
The bunch is thus focused by the $v_\mathrm{b}\times B_{\theta}$ term (see Section 1). %
This is true whatever the bunch charge sign. %
\item The fields have continuous longitudinal ($\sin$, $\cos$) and radial $\left(R(r)\right)$ variations. %
This means that different longitudinal and radial parts of the bunches are accelerated and focused differently by the wakefields. %
This leads to a broad final energy spread and to emittance growth, respectively. %
It also means that upon propagation the bunch and the wakefields will evolve self-consistently. %
This is, in general, not desirable for an accelerator. %
\item Since the wakefields are driven by the (drive) bunch, they are also tied to the bunch, \ie they travel with the same velocity as the drive bunch (in the absence of evolution of the bunch or change in plasma density). %
This (phase) velocity of the wakefields is therefore also close to the speed of light (for $\gamma\gg1$ and without significant evolution of the bunch). %
This is possible because in a cold plasma the Langmuir electrostatic wave has zero group velocity~\cite{ffchen}. %
\end{itemize}

\subsection{Bunch size}
The cold plasma responds collectively to perturbations, with the fastest time scale given by ${\sim}1/\omega_{\mathrm{pe}}$ and the smallest spatial scale given by ${\sim }c/\omega_{\mathrm{pe}}$. %
The bunch transverse and longitudinal sizes can be adjusted to maximize the effect of the bunch on the plasma electrons and thus also on the wakefield's amplitude. %

\subsubsection{Longitudinal size}
For PBPAs, the plasma wave of interest is the Langmuir, electrostatic plasma wave with natural frequency $\omega_{\mathrm{pe}}$. %
It is natural to think that a perturbation (\ie particle bunches with $n_\mathrm{b}\ll n_{\mathrm{e}0}$) in time and space shorter rather than longer than the previously described scales is most effective at driving the wave. %
We note here that a periodic perturbation with period at these time or spatial scales is also very effective in driving the wake. %
However, most of today's experiments use a single, short particle bunch (or laser pulse). %
Interest in using periodic excitation was in vogue when sufficiently intense short laser pulses did not exist~\cite{bib:modena} and is being re-examined, in particular, to take advantage the large amounts of energy stored in long proton bunches (PWFA)~\cite{kumar} and to ease the power requirements on laser systems (LWFA)~\cite{hooker}. %

Solutions to Eq.~(\ref{eqn:wave}) are well known (see, for example, Refs.~\cite{oscillator,oscillator2}). %
They indicate that for a finite time or impulse excitation and for weak damping cases, it is the amplitude of the Fourier component of the excitation at the system resonant frequency (here, $\omega_{\mathrm{pe}}$) and the damping factor that determine the oscillation amplitude. %
Most particle bunches have a density or current distribution that is close to Gaussian: $n_\mathrm{b}(t)\sim n_{\mathrm{b}0}\exp(-t^2/2\sigma_t^2)$ or  $n_\mathrm{b}(\xi)\sim n_{\mathrm{b}0}\exp(-\xi^2/2\sigma_{\xi}^2)$, where $\xi$ is the position along the bunch (sometimes $z$ is also used). %
The Fourier transform of such a bunch profile is simply $\tilde{n}_\mathrm{b}(k)=\tilde{n}_{\mathrm{b}0}\exp(-k_{\mathrm{pe}}^2\sigma_{\xi}^2/2)$. %
Therefore, when $k_{\mathrm{pe}}\sigma_{\xi}\le\sqrt{2}$, significant excitation occurs. %
For single-bunch experiments, where particles in the front and core of the bunch drive the wakefields and lose energy, while particles in the back gain energy from the wakefields, the bunch length $k_{\mathrm{pe}}\sigma_{\xi}\cong\sqrt{2}$ is optimum~\cite{lee00}. %

\begin{figure}[ht]
\begin{center}
\includegraphics[width=8cm]{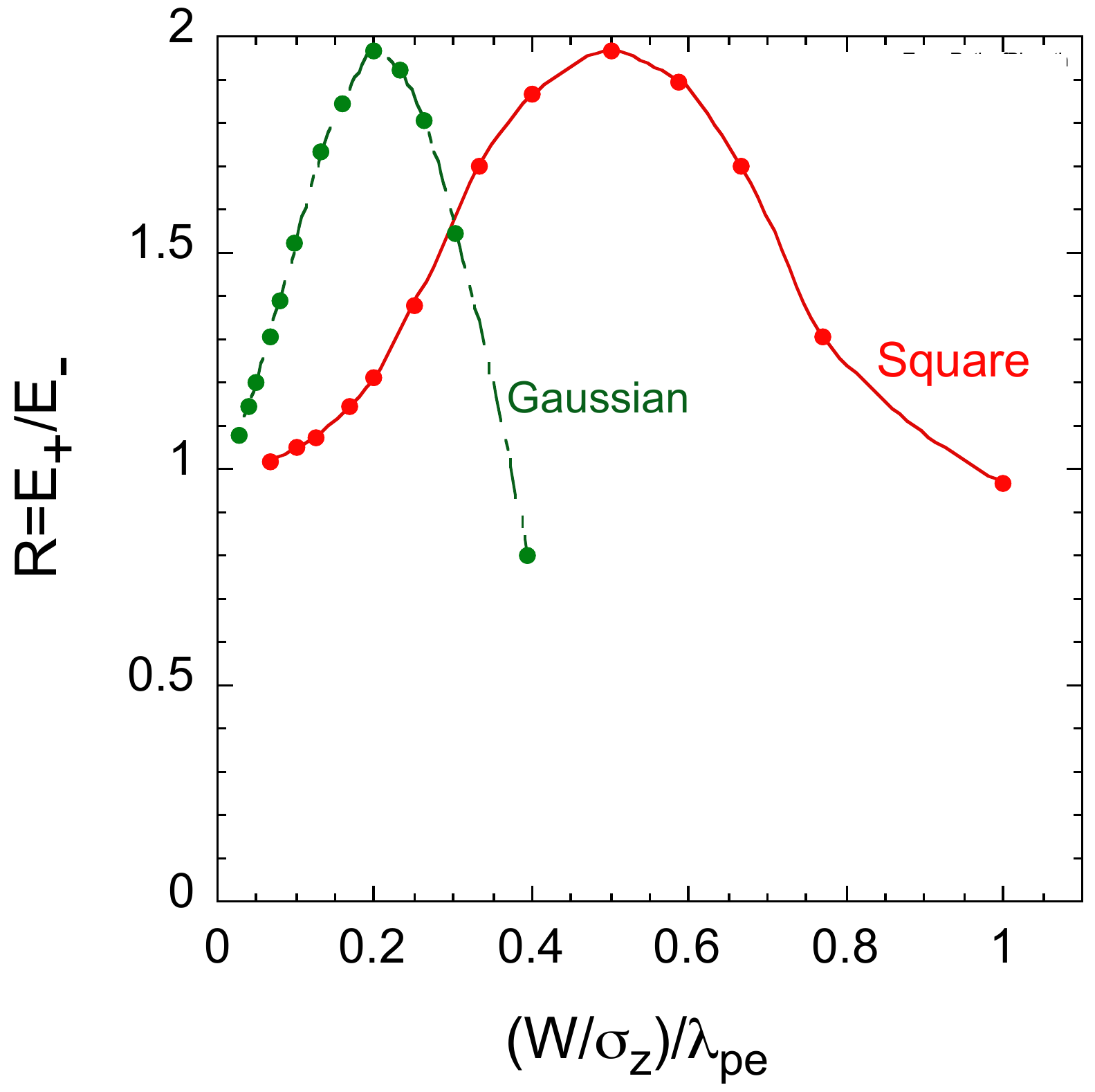}
\caption{Transformer ratio $R=E_{+}/E_{-}$ as a function of the width $W$ of a square longitudinal profile electron bunch or Gaussian r.m.s.\ width $\sigma_z $, both normalized to the plasma wavelength $\lambda_{\mathrm{pe}}$. }

\query{In Figure 3, please change $y$-axis label to $R=E_{+}/E_{-}$ and $x$-axis label to $(W, \sigma_z)/\lambda_{\mathrm{pe}}$. Please remove label `TransRatioofBlength'. Thank you.}

\label{fig:TransRatioofBlength}
\end{center}
\end{figure}

Using linear theory, one can calculate the wakefields driven by bunches with various parameters for a fixed plasma density. %
However, for a meaningful comparison, one needs to specify which parameters are held constant. %
For example, as the r.m.s.\ bunch length is varied, is the number of particles in the bunch or the bunch current kept constant? %
A way to get around this difficulty is to use the transformer ratio as a figure of merit instead, since wakefield amplitudes can always be obtained by increasing the charge (in linear theory). %
For example, one can use Eq.~(\ref{eq:MultiBunches_long1}) to calculate the wakefield's amplitude and then calculate $R=E_{+}/E_{-}$. %
Figure~\ref{fig:TransRatioofBlength} shows the transformer ratio obtained for various bunch lengths $W$ of a square bunch or the r.m.s.\ length $\sigma_z$ of a Gaussian bunch, with all other parameters kept constant. %
For the case of the square bunch, the maximum $R=2$ is obtained for $W=\lambda_{\mathrm{pe}}/2$. %
In this case, all bunch particles reside in the decelerating phase of the wakefields. %
For the Gaussian bunch case, the maximum $R=2$ is reached for $\sigma_z/\lambda_{\mathrm{pe}}\cong0.2$ or $k_{\mathrm{pe}}\sigma_z\cong1.25$. %
This is close to the predicted $k_{\mathrm{pe}}\sigma_z\cong\sqrt{2}$. %

This scaling of the wakefield with bunch length and its corresponding increase in amplitude~\cite{lee00} with plasma density has been observed in experiments with electron bunches. %
A gradient of the order of ${\sim}100\UMeV\Um^{-1}$ with bunches with $\sigma_z\sim600\Uum$ in a plasma with $n_\mathrm{e}\sim2\times10^{14}\Ucm^{-3}$ was measured~\cite{muggli04}. %
A gradient in excess of ${\sim}50\UGeV\Um^{-1}$ with bunches with $\sigma_z\sim20\Uum$ in a plasma with $n_\mathrm{e}\sim2.3\times10^{17}\Ucm^{-3}$ has also been measured~\cite{hogan05,blumenfeld}. %
The scaling has also been confirmed by numerical simulations~\cite{joshi02}. %

\subsubsection{Multiple drive bunches}
\begin{figure}[ht]
\begin{center}
\includegraphics[width=8cm]{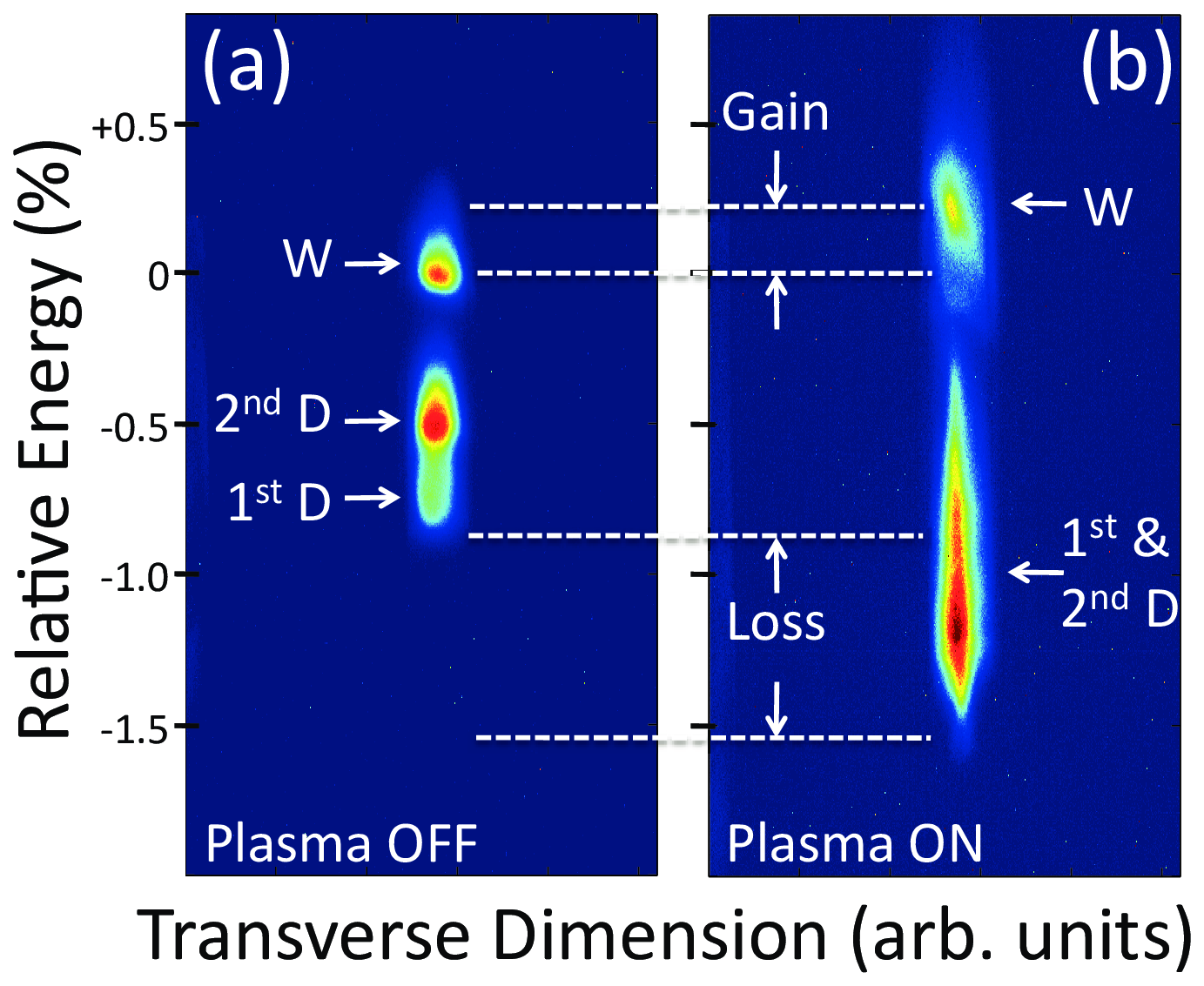}
\caption{Energy spectra obtained without (a) and with (b) plasma for the case of two drive bunches (labelled 1st and 2nd D) spaced longitudinally by $\Delta z=\lambda_{\mathrm{pe}}$ (at this plasma density) and a witness bunch (labelled W) following the second drive bunch at $1.5\lambda_{\mathrm{pe}}=1.5\Delta z$ (from Ref.~\cite{muggli09}). %
In panel (b), the two drive bunches lose energy to the plasma and the witness bunch gains energy from it. %
The two drive bunches appear merged on panel (a), owing to coherent synchrotron radiation (CSR) effects~\cite{bib:yakimenkocsr} but are separated in time~\cite{bib:muggli08, {bib:muggli10}}. %
}

\query{Please change hyphens in numbers in axis to proper minus signs. Thank you.}
\label{fig:TwoWandD}
\end{center}
\end{figure}

We have seen that Eq.~(\ref{eqn:wave}) has a natural periodic solution with period $2\pi/\omega_{\mathrm{pe}}$. %
We also saw that the plasma, as an oscillator with natural frequency $\omega_{\mathrm{pe}}/2\pi$ selects the frequency content of the drive bunch or train at that frequency. %
It is therefore also natural to use a (pre-formed) train of bunches with a separation equal to the plasma period to resonantly drive the wakefield. %
In linear theory, the wakefields driven by a train are simply the algebraic sum of the wakefields driven by each bunch (with the proper phase). %
Driving wakefields with multiple bunches has been demonstrated, for example, in Ref.~\cite{muggli09}. %
Figure~\ref{fig:TwoWandD} shows the case of two drive bunches and a witness bunch. %
The plasma density is chosen so that the plasma wavelength is equal to the bunch spacing ($\lambda_{\mathrm{pe}}(n_{\mathrm{e}0})=\Delta z$), so that both bunches are in the decelerating phase of the total wakefields. %
The witness bunch follows at a distance $1.5\lambda_{\mathrm{pe}}$ from the second drive bunch, is therefore in the energy gain phase of the wakefields and gains energy from the plasma.

There is little advantage in using a train, unless it is easier to generate a train than a single bunch, or the train is used either to increase the energy extraction efficiency~\cite{maeda} or to increase the transformer ratio~\cite{laziev}. %
In the former case, the drive and witness bunch are interleaved with a new drive bunch, to replenish the wakefields of the energy extracted by the previous witness bunch. %
The multi-bunch scheme, relevant for a high-efficiency collider, is not discussed here. %
In the latter case, the train is a particular form of shaped bunch. %

A transformer ratio greater than two with two bunches has been demonstrated in a dielectric based accelerating structure~\cite{jing}, showing the applicability of the wakefield principles to various acceleration schemes.

The case of driving wakefields with a self-generated train of bunches using the self-modulation instability is described in Section 7. %

\subsubsection{Transverse size}
The bunch transverse size must be kept smaller than the (cold) plasma skin depth $c/\omega_{\mathrm{pe}}$ or, equivalently, such that $k_{\mathrm{pe}}\sigma_{r0}<1$, where $k_{\mathrm{pe}}=\omega_{\mathrm{pe}}/c$. %
In the opposite case, the bunch is subject to the current filamentation instability (CFI)~\cite{cfi}.  The occurrence of CFI results in the breaking up of the bunch current density into filaments of larger current density at the $c/\omega_{\mathrm{pe}}$ scale in the transverse direction, as observed experimentally~\cite{allen}.

The ability to focus the beam to a small size often limits the maximum density, and thus the maximum accelerating gradient, at which the PBPA can be operated without risk of filamentation. %
This density is given by
\begin{equation}
n_\mathrm{e}\le\frac{1}{4\pi}\frac{1}{r_0\sigma_{r0}^2}\ .
\label{eqn:maxdens}
\end{equation}
Here, $r_0=\frac{1}{4\pi\epsilon_0}\frac{e^2}{m_0c^2}\cong2.82\time10^{-15}\Um$ is the classical radius of the electron. %

\subsection{Maximum accelerating field}
The linear PWFA theory is valid for small perturbations of the equilibrium quantities (\eg $n_1\ll n_{\mathrm{e}0}$). %
In this case the perturbations are sinusoidal. %
Although not strictly correct, one can estimate the maximum longitudinal electric field that can be expected by assuming a density perturbation with amplitude equal to the initial density: $n_1=n_{\mathrm{e}0}$. %
Using the Fourier transform of Poisson's equation with $\left|\nabla\right|\rightarrow k_{\mathrm{pe}}$,

\begin{equation}
\overrightarrow{\nabla}\cdot \overrightarrow{E}=\frac{\rho}{\epsilon_0}\rightarrow k_{\mathrm{pe}}E=\frac{\omega_{\mathrm{pe}}}{c}E=\frac{-en_{\mathrm{e}0}}{\epsilon_0}=\frac{e^2n_{\mathrm{e}0}}{m_\mathrm{e}\epsilon_0}\frac{m_\mathrm{e}}{(-e)}\rightarrow\frac{\omega_{\mathrm{pe}}}{c}\left|E\right|=\frac{m_\mathrm{e}\omega_{\mathrm{pe}}^2}{e}\ , \
\label{eqn:toEWB}
\end{equation}
and the corresponding field, known as the cold plasma wave-breaking field, $E_{\mathrm{WB}}$ is given by
\begin{equation}
E_{\mathrm{WB}}=\frac{m_ec\omega_{\mathrm{pe}}}{e} \ .
\label{eqn:EWB}
\end{equation}
(The ions are assumed not to move; their density remains uniform and thus they do not contribute to components at wave number $k_{\mathrm{pe}}$.) It is clear that non-linearities will appear long before the plasma density perturbation reaches $n_{\mathrm{e}0}$ and the field reaches $E_{\mathrm{WB}}$. %
The wakefields become non-sinusoidal (including higher harmonics) and particle trajectories cross~\cite{dawson}, invalidating the assumption of fluid behaviour for the plasma. %
However, this value is a good estimate of the field amplitudes that can be reached in a PBPA, as shown for example by simulations for the PWFA~\cite{joshi02}. %
Results in Ref.~\cite{joshi02} show that the wakefield amplitude and scaling generally follow those predicted by linear theory even into the non-linear regime. %
The expression for $E_{\mathrm{WB}}$ is often quoted in an engineering form,
\begin{equation}
E_{\mathrm{WB}}\propto0.96\UV[G]\Um^{-1}\sqrt{n_\mathrm{e}[10^{14}\Ucm^{-3}]},
\label{eqn:EWB-kg}
\end{equation}
and is often used to show the potential of PBPAs for high-gradient acceleration. %
Plasma electron densities of $10^{14}$ to $10^{19}\Ucm^{-3}$ are routinely used in PBPAs with corresponding accelerating fields in the range of a few $\UV[G]\Um^{-1}$ to hundreds of $\UV[G]\Um^{-1}$. %

Crossing of the plasma electron trajectories associated with reaching $E_{\mathrm{WB}}$~\cite{dawson} leads to plasma electron self-injection in the LWFA scheme. %
We note here that PWFAs usually operate at wakefield phase velocities too large and longitudinal field amplitudes too small to trap plasma electrons. %
However, ionization injection, now often used in the LWFA, was first observed in the PWFA~\cite{oz}. %

\subsection{Beam loading}

Beam loading is used in many accelerators, in particular to decrease the energy spread resulting from the finite length of bunch when compared with the accelerator wavelength and to maximize the energy extraction or transfer efficiency~\cite{beamloading}. %
It will naturally play an important role in PBPAs because they operate at high frequencies ($\omega_{\mathrm{pe}}/2\pi>100\UGHz$) and thus with very short wave periods (${<}1\Umm$) and accelerator cavity size (${<}1\Umm^3$). %

Beam loading can be seen as the simple addition of the bunch wakefields to the fields of the accelerating structure (RF structure or plasma wakefields of the drive bunch). %
The bunch cannot be placed at the peak of the field, where the accelerating field is constant for a narrow phase range, since (in the linear regime and in a PBPA) the transverse field becomes defocusing at that point. %
This is not the case in most RF structures that have no significant transverse field component (TM modes). %
It has to be placed ahead of the peak for effective beam loading where the field derivative is opposite to its own. %

One can illustrate beam loading by using the PWFA linear equations given earlier~\cite{katsou}. %

\begin{figure}[ht]
\begin{center}
\includegraphics[width=10cm]{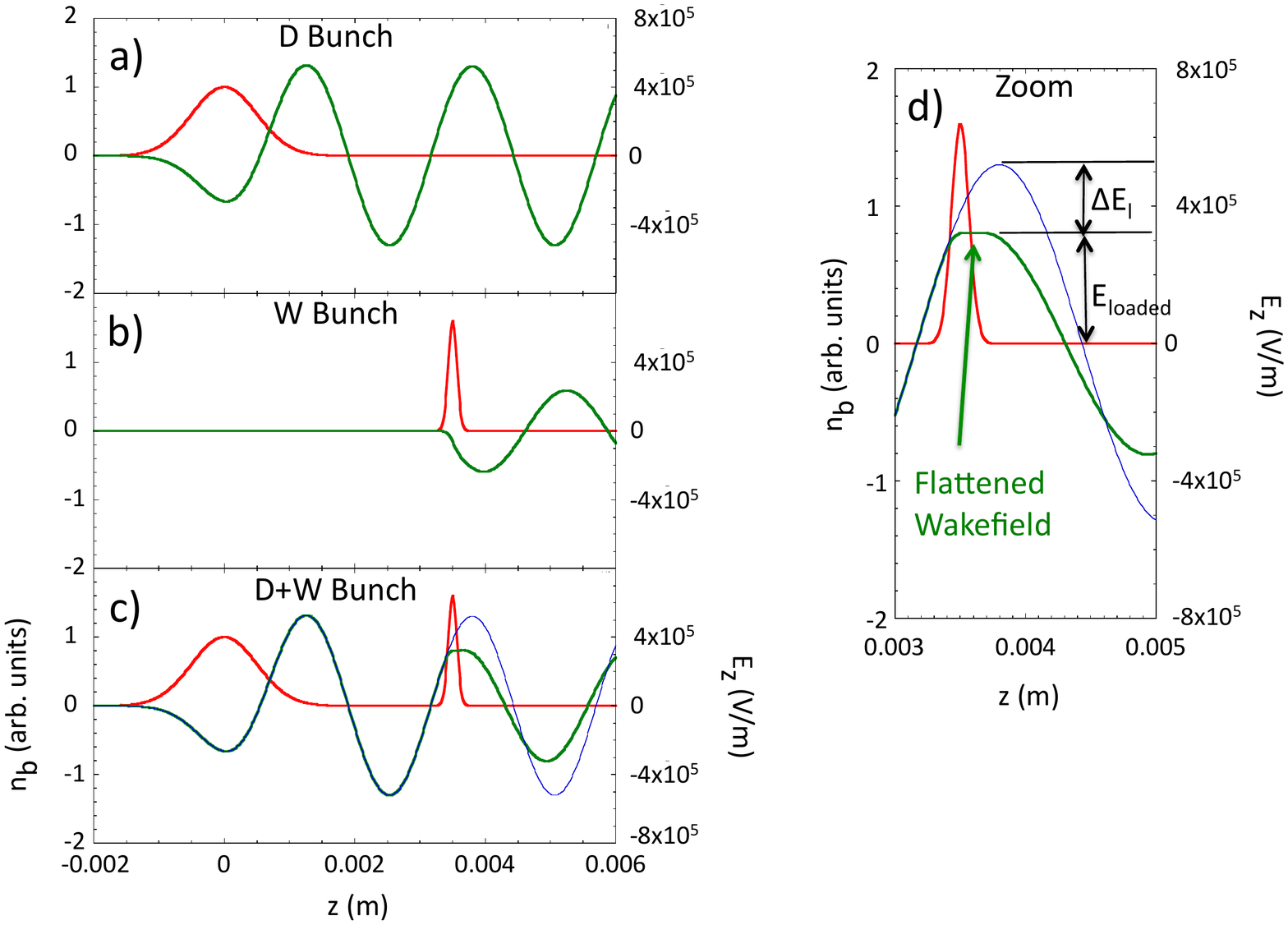}
\caption{Linear beam loading example: (a) drive bunch density profile (red line) and longitudinal wakefield $E_z$ (green line), (b) same for the witness bunch, (c) same for the drive and witness bunches together. %
The field of the drive bunch only is shown as the blue line in panel (c). %
A zoom around the witness bunch is shown in panel (d). %
The bunches move to the left.}

\query{In Figure 5, please correct maths: $n_{\mathrm{b}},z, E_z, E_{\mathrm{loaded}}, \Delta E_{\mathrm{l}}$. Please set numbers correctly, \eg $-4\times 10^{-5} $ Please note that variables are set italic but labels are set roman. Please remove illegible labels in top right corners. Thank you.}

\label{fig:BeamLoading}
\end{center}
\end{figure}

Figure~\ref{fig:BeamLoading}(a) shows a Gaussian bunch and the longitudinal wakefields $E_z$ it is driving in the plasma. %
Note that this bunch drives wakefields with a transformer ratio very close to two. %
Figure~\ref{fig:BeamLoading}(b) shows the wakefields driven by a shorter, following, witness bunch. %
This bunch is made shorter than the drive bunch, in order for it to sample a narrow phase range (${\ll}\pi/2$), to minimize the energy spread expected for the field variation along the bunch. %
The total field is shown in Fig.~\ref{fig:BeamLoading}(c). %
The parameters of the witness bunch (charge, length, relative position, \etc) were chosen to illustrate beam loading. %
The witness bunch is located in a region of increasing field so that the addition of its own field flattens the total field (Fig.~\ref{fig:BeamLoading}(d)). %
This more constant or flat region of accelerating field within the witness bunch reduces its final energy energy spread. %
However, this more uniform field ($E_{\mathrm{loaded}}$) is also lower than the peak field, by almost 25$\%$ (??) in this case
($\Delta E_\mathrm{l}$), decreasing the effective transformer ratio by the same fraction. %
The fraction of energy, proportional the field amplitudes squared, extracted from the wakefields by the witness is on the order of 50$\%$. %
The energy spread $\Delta E$ reduces from about 25$\%$ to less than 8$\%$ in this non-optimized case.%

\query{In the text `uniform field ($E_{\mathrm{loaded}}$) is also lower than the peak field, by almost 25$\%$ (??)', please sort out the question marks. Also, I am not sure what the ell in $\Delta E_\mathrm{l}$ stands for?}

Note that beam loading in the non-linear regime has also been studied~\cite{tzoufras, tzoufras2}. %
Narrow energy spread and high-energy transfer efficiency can be reached, at the expense of tailoring the witness bunch current profile. %
Beam-loading effects have been observed in recent PWFA experiments, leading to significant wakefield to witness bunch energy transfer efficiency and to narrow final energy spread~\cite{bib:litos14}. %

\section{Non-linear PWFA regime}
In the linear PWFA regime, the wakefields vary along and across the bunch and field structure (see Eqs.~(\ref{eq:MultiBunches_long1}) and (\ref{eq:MultiBunches_tran1})). %
This leads to significant final energy spread and to emittance growth. %
In the non-linear PWFA regime, the accelerating field in the pure ion column is independent of radius. 
It varies longitudinally, but this can be mitigated by using beam loading. %
The focusing field varies linearly with radius, which preserves the emittance of a bunch with Gaussian position and velocity distributions. %
It is also independent of the longitudinal position. %
Therefore, the non-linear PWFA regime has significant advantages over the linear regime, at least for electron beam (or negatively charged particle bunches). %

While there is no strict theory for the PWFA in the non-linear regime, an interesting theory based on the parameters of the electron bubble and of the plasma electron sheet that sustains it has been developed~\cite{lu, lu2}. %
It was used, for example, for the beam loading study in the non-linear regime mentioned previously~\cite{tzoufras,tzoufras2}. %
However, it is beyond the scope of this manuscript. %

\subsection{Electron bunch propagation}

In a vacuum, a (Gaussian) beam expands because of its emittance, with a characteristic increase in transverse size of $\sqrt{2}$ per $\beta_0$ propagation distance. %
This also means that the bunch density (${\propto}1/\sigma_r^2$) decreases by a factor of two, as does the wakefield amplitude that it can drive (see Eq.~(\ref{eq:MultiBunches_long1}) for the linear regime case). %

The plasma provides (transverse) focusing through the transverse wakefield $W_{\perp}$ (see Eq.~(\ref{eq:MultiBunches_tran1}) for the linear case), so that the beam can remain transversely small and keep driving large amplitude wakefields. %

In the linear regime, the beam continuously evolves in transverse size and distribution, since $W_{\perp}$ is a function of $(r,\xi)$ and its emittance grows. %
It can be shown that the r.m.s.\ transverse position and velocity distributions are preserved by a transverse force that varies linearly with radius: $F_{\perp}\sim r$. %
This is the case in the non-linear regime, in which $n_\mathrm{b}\ge n_{\mathrm{e}0}$ and the plasma structure sustaining the wakefields consists of a `bubble' that is void of plasma electrons. %
It thus consists of a pure ion column with uniform density (if there is no ion motion). %
In the ion column, there is no magnetic field since there is no plasma return current (it flows in a sheet around the bubble). %
Therefore, the transverse force experienced by the electrons in the column is simply $\overrightarrow{F}_{\perp}=-e(\overrightarrow{E}_r+\overrightarrow{v}_\mathrm{b}\times \overrightarrow{B}_{\theta})=-e\overrightarrow{E}_r$. %
The electric field of the ion column can be estimated using Poisson's equation, assuming that it is infinitely long and cylindrically symmetrical, with $n_{\mathrm{i}0}=n_{\mathrm{e}0}$:
\begin{equation}
\overrightarrow{\nabla}\cdot\overrightarrow{E}=\frac{\rho_{\mathrm{ion}}}{\epsilon_0}\rightarrow E_r=\frac{1}{2}\frac{n_{\mathrm{e}0}e}{m_\mathrm{e}}r\ .
\label{eqn:Eioncol}
\end{equation}
Thus, $F_{\perp}=eE_r\sim r$. %

Bunch particles have an equation of motion in the pure ion column given by
\begin{equation}
\frac{\mathrm{d}^2r(z)}{\mathrm{d}z^2}+K^2r(z)=0\ .
\label{eqn:partoscil}
\end{equation}
For the case of the pure ion column, $K^2=\frac{1}{\gamma m_\mathrm{e}c^2}\frac{F_{\perp}}{r}=\frac{1}{2\gamma m_\mathrm{e}c^2}\frac{n_\mathrm{e}e^2}{\epsilon_0}$. %
This is a harmonic oscillator equation with general solution $r(z)=r_0\exp\left(\pm \mathrm{i}k_{\upbeta}z\right)$, where $k_{\upbeta}=\frac{1}{\sqrt{2\gamma}}\left(\frac{n_\mathrm{e}e^2}{\epsilon_0m_\mathrm{e}c^2}\right)^{1/2}=\frac{k_{\mathrm{pe}}}{\sqrt{2\gamma}}$ is the betatron wave number. %

The evolution of the bunch transverse r.m.s.\ size is then described by the envelope equation (see Ferrario's lecture), which, neglecting acceleration, reads
\begin{equation}
\frac{\mathrm{d}^2\sigma_r(z)}{\mathrm{d}z^2}+K^2\sigma_r(z)=\frac{\epsilon^2_\mathrm{g}}{\sigma_r^3(z)}\rightarrow \sigma_r''+K^2\sigma_r=\frac{\epsilon^2_\mathrm{g}}{\sigma_r^3}\ .
\label{eqn:envelope}
\end{equation}
This equation is very similar to that for the individual particles (Eq.~(\ref{eqn:partoscil})), with $r$ replaced by $\sigma_r$ and the addition of the emittance term. %
It shows that at locations where $\sigma_r$ is large, the $K^2\sigma_r$ term dominates the $\epsilon_\mathrm{g}/\sigma_r^3$  term, $\sigma_r''<0$ and the beam is focused by the external force (assuming $K^2>0$). %
As the beam is focused, its size becomes smaller, the $\epsilon_\mathrm{g}/\sigma_r^3$ term eventually dominates the $K^2\sigma_r$ term, and the bunch diverges because of its emittance. %
Therefore, in general, the bunch transverse size oscillates between a minimum and maximum (\ie remains positive). %
The particles also oscillate but, of course, cross the axis. %
There is a situation in which the bunch size does not oscillate. %
Envelope oscillation amplitudes can be calculated if one considers the simple and practical case of a bunch focused at the entrance of the plasma (\ie $\sigma_r'(z=0)=0$) with size $\sigma_{r0}$. %
In this case, Eq.~(\ref{eqn:envelope}) can be integrated to obtain the smallest and largest bunch sizes as a function of the matched bunch size $\sigma_{r\mathrm{m}}$. %
Rewriting Eq.~(\ref{eqn:envelope}) as
\begin{equation}
\sigma_r''=\left(\frac{\epsilon^2_\mathrm{g}}{\sigma_r^4}-K^2\right)\sigma_r\ ,
\label{eqn:envelope2}
\end{equation}
and first setting the term in parentheses to zero and with the initial conditions specified leads to
\begin{equation}
\sigma_{r\mathrm{m}}^4=\frac{2\epsilon_0\gamma m_\mathrm{e}c^2\epsilon_\mathrm{g}^2}{n_{\mathrm{e}0}e^2}\ ,
\label{eqn:sigmam}
\end{equation}
or the more general condition for the bunch and plasma parameters:
\begin{equation}
\frac{\sigma_{r\mathrm{m}}^4n_\mathrm{e}}{\gamma\epsilon_\mathrm{g}^2} =\frac{2\epsilon_0m_\mathrm{e}c^2}{e^2}\ .
\label{eqn:sigmam2}
\end{equation}
When the conditions: $\sigma_r=\sigma_{r0}$, $\mathrm{d}^2\sigma_r/\mathrm{d}z^2=\mathrm{d}\sigma_r/\mathrm{d}z=0$ are satisfied at the plasma entrance ($z=0$), the beam radius remains constant along the plasma and the beam is said to be matched to (the focusing force of) the plasma. %

Note that $\sigma_{r\mathrm{m}}\sim\gamma^{1/4}$ is weakly dependent on particle energy, which may allow for adiabatic matching of the bunch size to the plasma focusing upon acceleration. %
Note also that while the bunch envelope size does not oscillate in the matched case, particles do. %

In general, the (unmatched) beam envelope size oscillates in the ion column; it reaches the two sizes $\sigma_1=\sigma_0$ and $\sigma_2=\frac{\sigma_{r\mathrm{m}}}{\sigma_0}\sigma_{r\mathrm{m}}$. %
These sizes can be calculated by multiplying Eq.~(\ref{eqn:envelope2}) by $\sigma_r'$, integrating it and setting $\sigma_r'=0$. %
Therefore, when $\sigma_0>\sigma_{r\mathrm{m}}$, $\sigma_2<\sigma_{r\mathrm{m}}$, and when $\sigma_0<\sigma_{r\mathrm{m}}$, $\sigma_2>\sigma_{r\mathrm{m}}$. %
This means that the maximum beam size is \emph{always} larger than or equal to the matched size. %

The electrons oscillate with the betatron wavelength $\lambda_{\upbeta}=\sqrt{2\gamma}\lambda_{\mathrm{pe}}$, and the bunch envelope with a periodicity half of $\lambda_{\upbeta}$. %
The beam envelope oscillations were clearly observed with an electron beam by changing the plasma density~\cite{clayton}. %
Matching of the bunch to the plasma was also observed~\cite{muggli04}. %

The bunch electrons oscillating in the ion column emit synchrotron radiation called betatron radiation (see lecture by K. Ta Phuoc). %
This radiation was observed for the first time in a PBPA as X-radiation \cite{wang}. %
It was later observed as $\gamma$-radiation at larger plasma densities~\cite{johnson}. %
This is now a main feature of the LWFA, which is a very interesting source of (betatron) radiation. %
These considerations on the bunch transverse size oscillation indicate that minimum betatron power is emitted when the bunch is matched to the plasma, since the betatron radiation power scales as $P_{\mathrm{betatron}}\sim r_0^2\sim\sigma_{r, \mathrm{max}}^2\ge\sigma_{r, \mathrm{m}}^2$. %
This can be used experimentally to find the matching condition by minimizing the amount of betatron radiation emitted by the bunch. %

\begin{figure}[ht]
\begin{center}
\includegraphics[width=8cm]{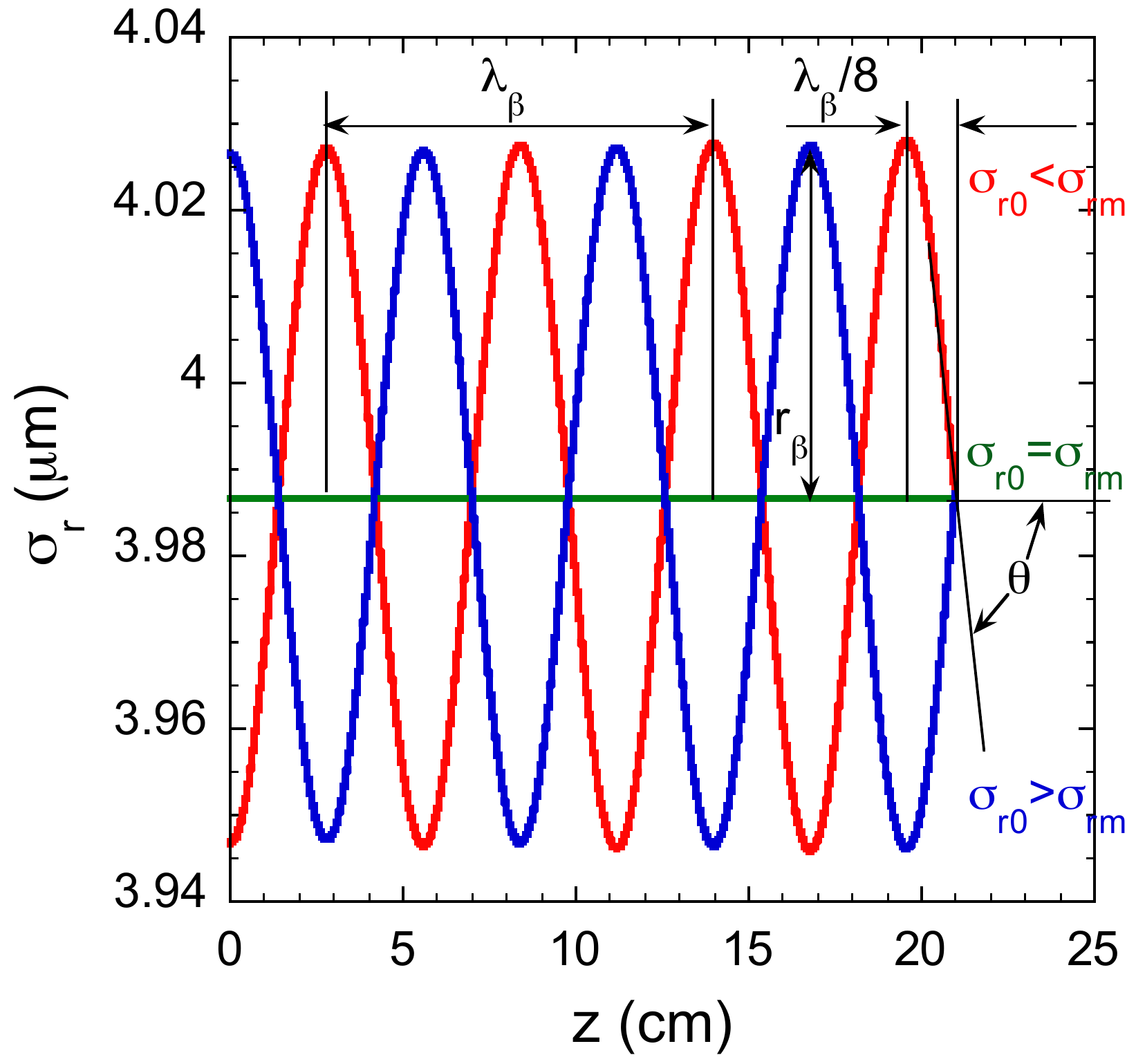}
\caption{Bunch envelope size along a plasma with uniform density for the case of a matched bunch (green line) and a slightly mismatched bunch with initial size too large (blue line) or too small (red line). %
The betatron wavelength and exit angle are shown. %
}
\query{In Figure 6, please correct the maths: $z, \sigma_r, \lambda, r_\upbeta, \theta$,  \etc\ Note that variables (such as radial distance) are set in italic font but labels (such as `matched' or `betatron') are set in roman font. Please remove label `MatchedBeam' from top right corner.}
\label{fig:MatchedBeam}
\end{center}
\end{figure}

Matching the beam to the pure ion column focusing also has another advantage: it minimizes the sensitivity of the beam angle at the plasma exit as a function of the beam and plasma parameters. %
When matched, $\sigma_r''=0$ or $(\sigma_r')'=0$, \ie the exit angle is minimum. %
Figure~\ref{fig:MatchedBeam} shows the beam angle at the plasma exit as a function of the relative variation of the beam size at the plasma entrance and of the plasma density. %
This figure was obtained by integrating numerically the envelope Eq.~(\ref{eqn:envelope}) in a plasma with constant density (\ie $K^2=\mathrm{constant}$). %
When the bunch is not matched, the angle of the beam at the plasma exit is $\theta\cong r_{\upbeta}/(\lambda_{\upbeta}/8)$, where $r_{\upbeta}$ is the radial oscillation amplitude of the beam envelope around the matched radius size. %

\begin{figure}[ht]
\begin{center}
\includegraphics[width=8cm]{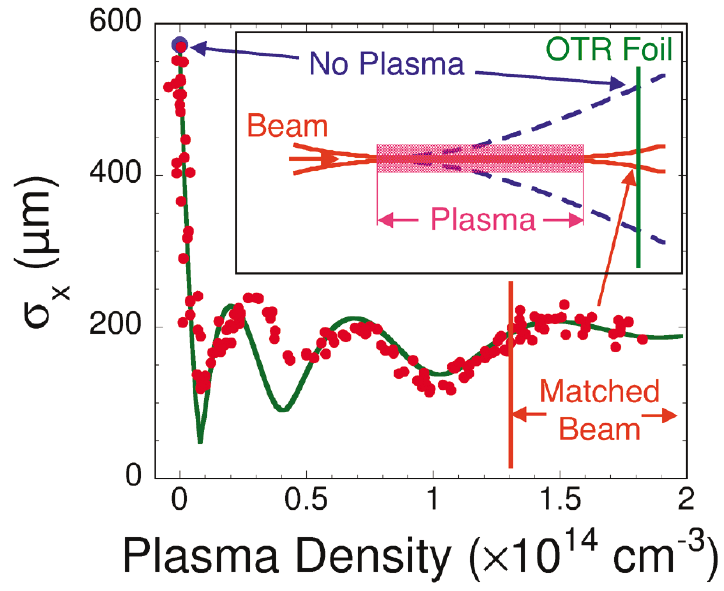}
\caption{Transverse bunch size measured as a function of plasma density a distance ${\approx}1\Um$ downstream from the plasma exit. %
 As the plasma density and the ion focusing column strength also increases (${\sim} n_{\mathrm{e}0}$), the beam approaches the matching condition observed in the range (1.25--2.5)$\times10^{14}\Ucm^{-3}$. %
See Ref.~\cite{muggli04} for experimental parameters. %
}
\query{Please correct $y$-axis label to $\sigma_x$ (\Uum) and $x$-axis label to have a proper minus sign instead of a hyphen. Thank you.}
\label{fig:SLACmatchedbeam}
\end{center}
\end{figure}

Matching of the beam to the focusing field of non-linear wakefields has been demonstrated experimentally. %
Figure~\ref{fig:SLACmatchedbeam}, from Ref.~\cite{muggli04}, shows the SLAC electron bunch transverse size measured downstream from the plasma, as a function of plasma density. %
Near the matched density, the variation in transverse size decreases, compared with the unmatched cases (lower density in this case). %

Matching was also demonstrated with lower bunch energy~\cite{bib:barov98}. %

\begin{figure}[ht]
\begin{center}
\includegraphics[width=13cm]{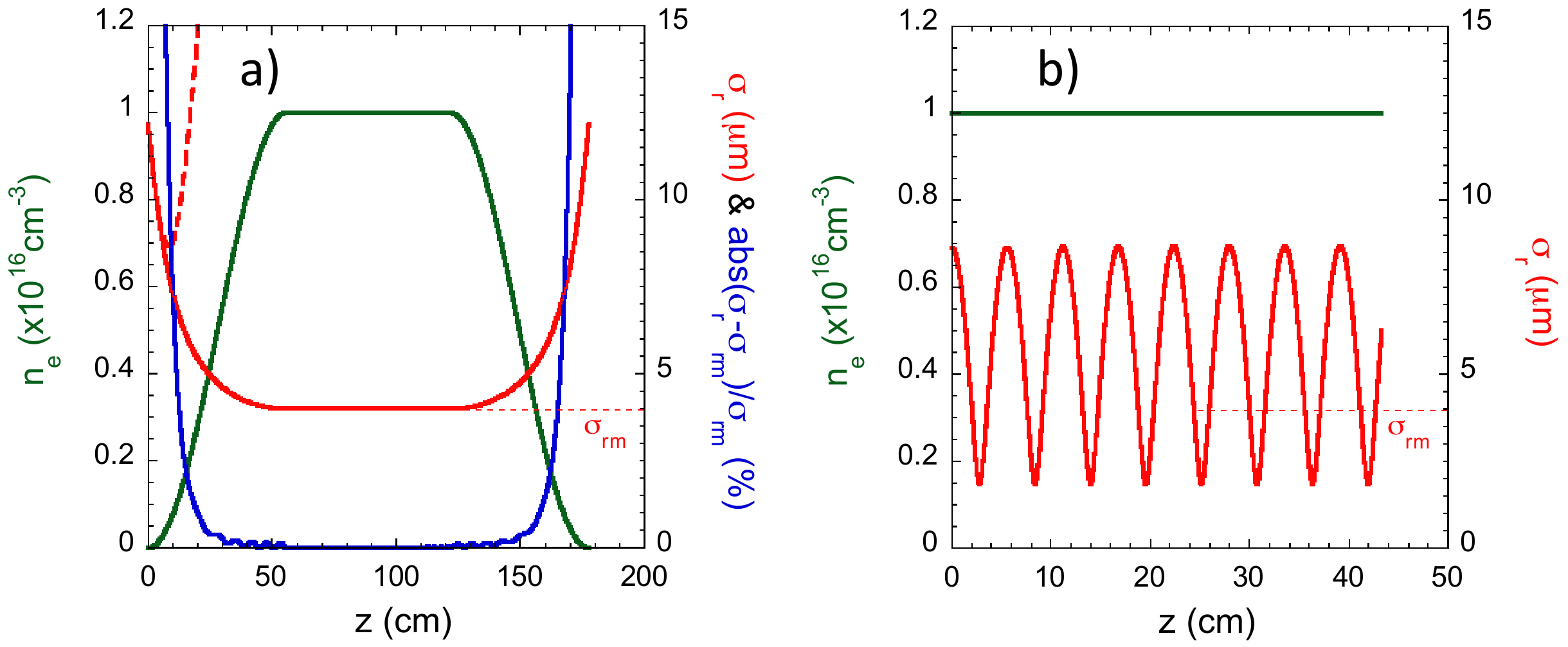}
\caption{(a) Beam transverse size in vacuum (dashed red line) and plasma (red line) with density ramps at the entrance and exit (green line). %
In this case, the beam vacuum focus is $\sigma_{r0}=8.6\Uum$ at $z=8.18\Ucm$, so that with the initial ramp, the beam reaches the matched radius $\sigma_{r\mathrm{m}}=3.8\Uum$ in the constant density region ($55.6<z<122\Ucm$). %
The total plasma length is $1.77\Um$ with a constant density of $10^{16}\Ucm^{-3}$, the particle's $\gamma$ is 56\,000, the beam normalized emittance is $5\times10^{-5}\Um\Urad^{-1}$ and the betatron wavelength in the constant density region is $\lambda_{\upbeta}=2\pi/K=11.2\Ucm$. %
The blue line represents the absolute value of the relative deviation between the beam size ($\sigma_r$) and the matched beam size (see Eq.~(\ref{eqn:sigmam})): $\left|\frac{\sigma_r-\sigma_{r\mathrm{m}}}{\sigma_{r\mathrm{m}}}\right|$. %
(b) Transverse beam size over the first $45\Ucm$ of plasma with the same density (and beam parameters) as panel (a) in the case where the beam is focused to the same $\sigma_{r0}=8.6\Uum$, but at the entrance of the plasma with a step-function density. %
The beam is mismatched and the envelope oscillations are indeed between the incoming size and that predicted by $\sigma_2\le\sigma_{r\mathrm{m}}$ in the previous paragraph, $\cong1.8\Uum$, in this case. %
}
\query{Please correct the labels: $n_{\mathrm{e}}(\times10^{16}\,\Ucm^{-3})$, $z$ (cm), $\sigma_r$ ($\upmu$m), $\mathrm{abs}(\sigma_r-\sigma_{r\mathrm{m}})/\sigma_{r\mathrm{m}}$  (\%), $\sigma_{r\mathrm{m}} $ and remove the small text labels at the top of the panels. Thank you.}
\label{fig:PlasmaRamp}
\end{center}
\end{figure}

The very strong plasma focusing force implies that the matched beam size is quite small. %
For example, for a typical SLAC electron beam with $\gamma\cong56\,000$ ($E_0\cong28\UGeV$) and a normalized emittance $\epsilon_\mathrm{N}=5\times10^{-5}\Umm\Umrad$, Eq.~(\ref{eqn:sigmam}) indicates that $\sigma_{r\mathrm{m}}\cong4\Uum$. %
This is a quite difficult size to produce at the plasma entrance. %
However, a continuous ramp in the plasma density, typical of real plasmas, rather than a step function, can be used to match the beam adiabatically to the plasma focusing force~\cite{marsh}. %
Adiabatic matching occurs when the variation of the bunch envelope size is small in one betatron period. %
Numerical integration of Eq.~(\ref{eqn:envelope2}) shows that this condition is easily satisfied. %
Plasma sources used for PWFA experiments, such as metal vapour sources~\cite{muggli99,li14}, naturally provide a neutral density ramp, which, once ionized by a laser pulse or by the bunch itself~\cite{oconnell06}, becomes a continuous plasma ramp. %
Figure~\ref{fig:PlasmaRamp} shows an example of a calculation in which the entrance plasma ramp is used to match the beam to the plasma and the exit ramp, to increase the bunch size and decrease its divergence in vacuum. %
Increasing the bunch size at the plasma exit decreases the emittance term, owing to the bunch energy spread (see M. Ferarrio's lecture). %

Matching into and out of the plasma is a very interesting and important research topic. %
Indeed, the larger focusing strength of the plasma, compared with that of conventional magnetic optics, leads to much shorter beta-functions in the plasma than outside of it, where the beam is used or conditioned to be injected into the next PBPA. %
Indeed with the definition of the beam beta (Eq.~(\ref{eqn:envelope2})) can be rewritten as
\begin{equation}
\beta_{\mathrm{matched}}=\frac{\sigma_{r\mathrm{m}}^2}{\epsilon_\mathrm{g}} =\frac{1}{K}\ .
\label{eqn:sigmam3}
\end{equation}

\query{There is no beta in Equation (23)?  Please clarify.}

It is important to note that the pure ion column, so favourable for an electron bunch or more generally for a negatively charged bunch, does not exist for a positively charged bunch (e$^+$, p$^+$, \etc). %
The plasma is asymmetrical for the charge signs of the two bunches since, in any case, only the light plasma electrons move (at the $1/\omega_{\mathrm{pe}}$ time scale). %
This difference has been observed with a positron bunch in conditions similar to those for an electron bunch~\cite{blue,muggli08} (see Section~\ref{PosAcc}). %
Focusing of a positron bunch by a short plasma has also been demonstrated~\cite{Ng}. %

\section{Plasma and magnetic focusing}

It is interesting to compare the focusing strength of the plasma ion column with that of a conventional quadrupole magnet. %
In a quadrupole magnet, focusing is achieved by imposing a magnetic field perpendicular to the particles' trajectories. %
The radius of curvature of the relativistic particle's trajectory in the magnetic field $B_{\perp}$ is given by the Larmor radius; $r_\mathrm{L}=\beta\gamma m_\mathrm{e}c/qB_{\perp}\cong\gamma m_\mathrm{e}c/qB_{\perp}$ ($\beta=(1-1/\gamma^2)^{1/2}\cong 1$). %
In a magnet (and magnetic field) of length $L$, this results in a deflection angle $\theta\ll1$, such that $r_\mathrm{L}\theta\cong L$. %
Therefore, $\theta\cong L/r_\mathrm{L}=qB_{\perp}L/\gamma m_\mathrm{e}c$. %
For the magnet to act as a lens, free of geometric aberration, the deflection angle and thus the magnetic strength must increase linearly with radius and reverse at the axis. %
This is the characteristic of the field with a quadrupole symmetry. %
A quadrupole magnet with a larger field gradient $B_{\perp}/r$ corresponds to a stronger, shorter focal length, focusing element. %
Quadrupoles usually use electromagnets to generate the field, but the largest field gradients are achieved in permanent magnet quadrupoles (PMQs). %
For example, the field gradient reached ${\sim}290\UT\Um^{-1}$ over $10\Ucm,$ with an aperture of $7\Umm$ in the PMQ of Ref.~\cite{pmq}. %

Using Eq.~(\ref{eqn:Eioncol}) for the ion column field, and dividing it by $r$ to obtain the gradient and by $c$ to produce the proper units of magnetic field, we obtain $E_r/rc\cong3\,kT/m$ for $n_\mathrm{e}=10^{14}\Ucm^{-3}$, as in Refs.~\cite{wang,clayton}. %
For the higher plasma densities of Refs.~\cite{johnson,hogan05,blumenfeld}, $n_\mathrm{e}=10^{17}\Ucm^{-3}$, one obtains $E_r/rc\cong3\,MT/m$. %
These values clearly show the potential for very strong focusing by plasmas in a scheme known as the plasma lens~\cite{plasmalens,chenfoc}. %
We note, however, that the transverse extent of the focusing region is limited to $\approx c/\omega_{\mathrm{pe}}<1\Umm$.

Since plasma lenses have such strong focusing gradients they can be shorter (and smaller) than magnetic ones. %
They could be of potential interest for replacing the km-long final focusing system of a collider~\cite{chen87}. %
This is another area where plasmas could contribute to reducing the size and cost of a future collider. %
We note however that plasma lenses have their own challenges, including the possible need of a \emph{drive bunch} to create the lens out of a neutral plasma. %

Quadrupoles are, by definition, focusing in one plane and defocusing in the other. %
They must therefore be used in pairs (a doublet) or in threes (a triplet) to build an element that focuses in both planes. %
They are also usually short compared with their focal length and are thin lenses (though they may be long magnets). %
A lens is said to be thin when its thickness or length is much shorter than its focal length. %

By contrast, the plasma ion column (or the plasma in general) is focusing (or defocusing) in both planes at the same time. %
Because of their large focusing strength, plasma lenses can be thin or thick lenses. %
Mismatched plasma accelerating sections are very thick focusing elements. %
The beam has multiple foci along the plasma, since its length can be many betatron wavelengths long (first focus near $\lambda_{\upbeta}/4$). %
For example, in the experiments described in Refs.~\cite{clayton,muggli04}, the beam has up to three foci along the plasma with $\lambda_{\upbeta}=81\Ucm$, maximum density $1.9\times10^{14}\Ucm^{-3}$, and length $\cong1.5\Um\cong2\lambda_{\upbeta}$. %
In the case of Ref.~\cite{bib:blumenfeld10}, the plasma length is about $46\lambda_{\upbeta}$. %

\subsection{Electron acceleration}

\begin{figure}[ht]
\begin{center}
\includegraphics[width=8cm]{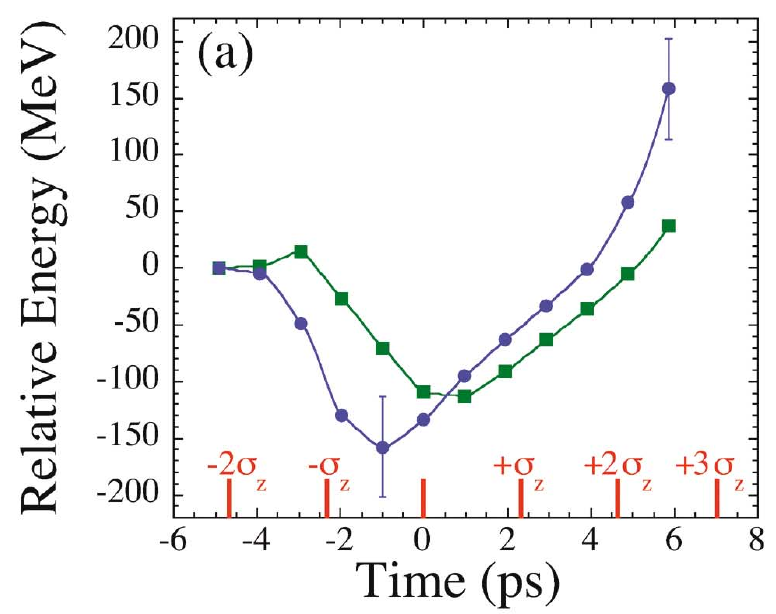}
\caption{Energy change, measured along the electron bunch with ${\sim}1\Ups$ resolution. %
The energy change reflects the shape of the wakefields along the bunch (the wakefield amplitude is the change in energy divided by the plasma length). %
The wakefields have the expected shape, \ie essentially zero ahead of the bunch (${\sim}-6\Ups$), with maximum energy loss in the core of the bunch (${\sim}0\Ups$) and, eventually, energy gain in the back of the single bunch (${>}+4\Ups$). %
The green line was obtained at a lower plasma density than the blue one. %
Energy gain in the back of the bunch (right-hand panel) is observed at a density for which $k_{\mathrm{pe}}\sigma_z\cong\sqrt{2}$. %
For parameters, see Ref.~\cite{muggli04}. %
}
\query{Please make the sigmas sloping, the subscript z italic (distance along the $z$-axis) and the minus signs proper minus signs instead of hyphens. Thank you.}

\label{fig:NonlinearWakefields}
\end{center}
\end{figure}

The accelerating field amplitude is usually not directly measured. %
The accelerating gradient is obtained by dividing the change in energy (loss or gain) by the (measured or assumed) plasma length. %
This is a quantity that is integrated over the plasma length, including the density ramp, in experiments. %
The measured energy gain and loss depend on the actual accelerating field along the plasma and on the fact that there are (or are not) particles sampling the field. %
The dependency of the longitudinal wakefield along the bunch was measured by time resolving, at the $ps$ scale, the energy change of the electrons along a single bunch~\cite{muggli04}. %
The general longitudinal shape of the field is as expected from simple consideration: decelerating, switching to accelerating (Fig.~\ref{fig:NonlinearWakefields}). %
Acceleration by trailing particles has been observed at a plasma density roughly satisfying $k_{\mathrm{pe}}\sigma_z\cong\sqrt{2}$ ($\sigma_z\cong700\Uum$ in this case). %

The transverse wakefield structure was also inferred~\cite{bib:yakimenko03} using the transverse and longitudinal wakefields $90^\circ$ out of the phase relation predicted by linear theory. %
Once again, the general agreement between experiment and theory is very good. %

\begin{figure}[ht]
\begin{center}
\includegraphics[width=12cm]{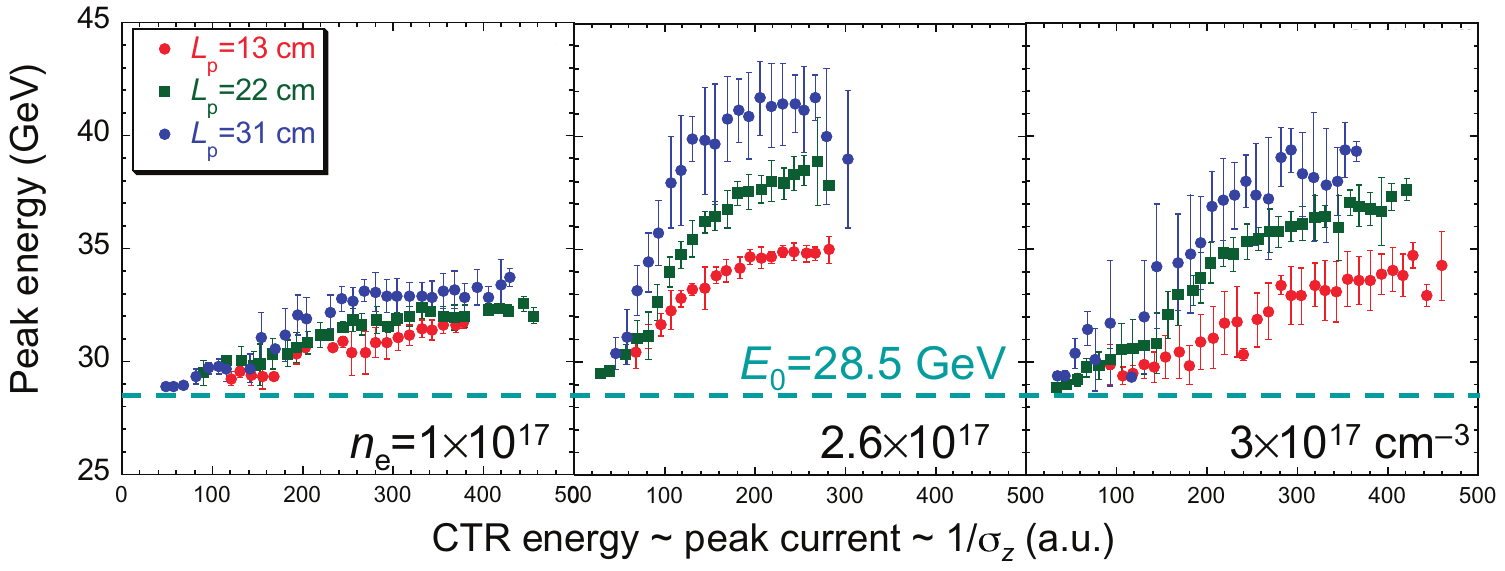}
\caption{Maximum energy of the particles (in a single bunch) measured at three plasma densities ($n_\mathrm{e}=1$, $2.6$ and $3\times10^{17}\Ucm^{-3}$) and each at three plasma lengths ($L_ \mathrm{p}=13$ (red symbols), $22$ (green symbols) and $31\Ucm$ (blue symbols) ) as the bunch is made shorter, and therefore the coherent transition radiation (CTR) energy emitted by the bunch increases. %
Higher CTR energy corresponds to higher peak current, shorter bunches. %
The error bars represent the r.m.s.\ energy measurement variation in each bin. %
For parameters, see Ref.~\cite{muggliNJP}. %
}

\query{Please change to $\sigma_z$. The other labels are fine (although the equals signs and multiplication signs could use a little space either side, if you have time). Thank you.}
\label{fig:Scaling}
\end{center}
\end{figure}

The maximum accelerating field (averaged over the plasma length) has been measured by varying the bunch length over a relatively wide range of plasma densities and lengths \cite{muggliNJP}. %
The results are shown in Fig.~\ref{fig:Scaling} from Ref.~\cite{muggliNJP}. %
In this case, $\sigma_z\cong20\Uum$ and the plasma density is much larger (${\sim}1000$ times larger than in~\cite{muggli04}) to approach $k_{\mathrm{pe}}\sigma_z\cong\sqrt{2}$. %
The left-hand panel of Fig.~\ref{fig:Scaling} shows that, at the lower density, the energy gains (and thus accelerating fields) are low because the bunch is too short $k_{\mathrm{pe}}\sigma_z>\sqrt{2}$ for this density. %
The bunch is also too short to sample the peak field. %
However, the energy gain increases with decreasing bunch length and increasing plasma length, as expected. %
At what turned out to be the optimum density (middle panel), the energy gain exhibits a (shallow) maximum for parameters close to $k_{\mathrm{pe}}\sigma_z\cong\sqrt{2}$ for the three plasma lengths. %
Wakefields are again probably larger with shorter bunch lengths, but there may be no particles to sample them. %
At the highest density (right-hand panel), the energy gain is lower than at the optimum density and increases with shorter bunches because more of the bunch charge participates in driving the wakefields as $k_{\mathrm{pe}}\sigma_z$ approaches $\sqrt{2}$ with decreasing $\sigma_z$. %
Since the bunch is long when compared with $\sigma_z$, there are probably particles to sample the maximum field amplitude. %

In Ref.~\cite{muggli04}, the measured (average) accelerating gradient is ${\sim}200\UMeV\Um^{-1}$, lower than expected from $E_{\mathrm{WB}}=1.3\UV[G]\Um^{-1}$ for $n_{\mathrm{e}0}=1.9\times10^{14}\Ucm^{-3}$. %
In Ref.~\cite{blumenfeld}, it is ${\sim}52\UGeV\Um^{-1}$, comparable to the expected value of $E_{\mathrm{WB}}=46\UV[G]\Um^{-1}$ for $n_{\mathrm{e}0}=2.3\times10^{17}\Ucm^{-3}$. %
Both experiments were in the $n_\mathrm{b}>n_{\mathrm{e}0}$ non-linear regime and used a single electron bunch to drive and sample wakefields. %
These results show the main features expected from the wakefield excitation. %
The results are in general agreement with the predictions of linear theory. %

Further experiments using two electron bunches showed a narrow energy spread of the witness bunch as well as good transfer efficiency between the wakefields and the witness bunch, due to a significant beam-loading effect~\cite{bib:litos14}. %

Measured energy gains are usually in excellent agreement with numerical simulation results with experimental parameters. %

\subsection{Positron bunch propagation}

In the linear PWFA regime, the wakefields are similar for an electron and a positron bunch with a simple phase change (see Figs.~\ref{fig:PWFAscematic1} and \ref{fig:PWFAscematic2}). %
Therefore, the accelerated bunch suffers from similar large final energy spread and emittance growth. %
In the non-linear regime, the locations of the wakefields that are focusing for a positron bunch are confined to the regions of return of the plasma electrons to the axis (see Figs.~\ref{fig:PWFAscematic1} and \ref{fig:PWFAscematic2}). %
These regions are small and have no strong variations of the fields. %
The locations that are accelerating are those immediately behind those pinch regions (regions that are decelerating for an electron bunch!). %
There is, therefore (and a priori), no good regime for accelerating a positron bunch to high energies while maintaining its incoming emittance. %

The propagation of positron bunches in plasmas has been studied experimentally. %
In particular, the focusing or plasma lens effect was observed in a short plasma~\cite{Ng}. %

\begin{figure}[ht]
\begin{center}
\includegraphics[width=10cm]{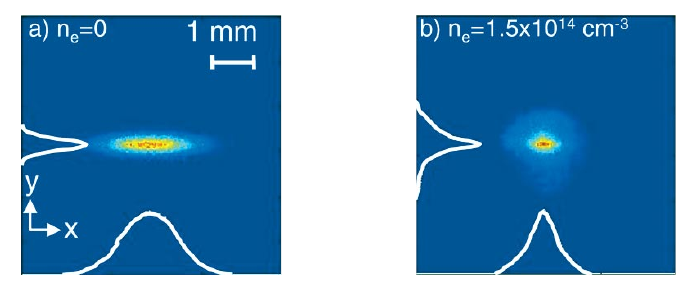}
\caption{Images of the positron transverse size obtained (a) without  and (b) with plasma, a distance ${\sim}1\Um$ downstream from the plasma exit. %
The images are obtained using optical transition radiation (OTR). %
The beam has an elliptical transverse shape without plasma (a) because it is focused to a round spot at the plasma entrance and has different emittances in the $x$- and $y$-planes (typical of linear colliders). %
The image with plasma (b) shows a focused core surrounded by a halo of particles. %
The halo is the result of the evolving and non-linear character of the plasma focusing force acting on the positron bunch. %
For parameters, see Ref.~\cite{muggli08}. %
}
\query{Please change the x, y and n  to italic. Thank you.}
\label{fig:PosFoc}
\end{center}
\end{figure}

Propagation in a long plasmas ($L\gg\lambda_{\mathrm{pe}}$) was also studied~\cite{hogan03,muggli08}. %
The difference for electron bunch propagation and the formation of a beam halo due to the non-linear focusing force was observed directly (Fig.~\ref{fig:PosFoc}). %
The associated emittance growth was inferred from numerical simulations that described the experiment~\cite{muggli08}. %

\subsection{Positron acceleration}\label{PosAcc}

The acceleration of positrons in plasmas has been observed~\cite{blue}. %
As expected, the measured energy change was lower than with an electron bunch with comparable parameters. %
The inferred wakefield along the bunch also has a more linear shape, \ie a greater sinusoidal dependency than that of the electron bunch~\cite{muggli04}. %

Recent results obtained at SLAC suggest that in some cases a situation favourable for large acceleration without significant emittance growth might exist in a uniform plasma. %

\subsection{Hollow plasma channel for positron bunches}

One possible solution to the emittance preservation during the acceleration of a positron bunch in a plasma is to use a hollow plasma channel. 
In such a channel, the plasma density is zero from the axis to a radius $r_\mathrm{c}$ and a constant density beyond $r_\mathrm{c}$. %
Simulation results show that $r_\mathrm{c}$ of the order of $c/\omega_{\mathrm{pe}}$ may be optimal in terms of accelerating gradient~\cite{lee01}. %
A drive bunch (electron or positron) displaces the plasma electrons in much the same way as in a uniform plasma and the plasma electrons converge towards the axis of the beam and channel. %
They sustain a longitudinal wakefield structure similar to that formed in a uniform plasma that has accelerating and decelerating regions. %
However, since there are no plasma ions in the channel, there are essentially no transverse forces to defocus a positron witness bunch. %
In a hollow plasma channel, a positron witness bunch can therefore be accelerated without the emittance growth associated with the unfavourable transverse wakefields. %
Since there are no focusing fields, the length of the acceleration is limited by the beta-function of the beam, which increases in size transversely and eventually reaches the channel radius. %
However, as previously noted, the emittance of the beams foreseen for a future accelerator might be low enough to allow for acceleration in metre-long plasma channels. %

Hollow plasma channels could be produced, for example, by heating a plasma with a long (${\sim}\Ups$) laser pulse and letting it expand~\cite{kimura} or by using specially designed phase optics to ionize a gas only in a hollow cylinder region~\cite{gessner}. %

\section{Quasi-linear PWFA regime}

An interesting regime, between the linear and non-linear regime can be reached and may prove advantagous for operation with multiple drive bunches. %
This regime is called the quasi-linear or weakly non-linear regime. 

This regime is reached when $n_\mathrm{b}\cong n_{\mathrm{e}0}$. %
In this case, blow-out is reached essentially only over the bunch volume, unlike in the non-linear regime, where it is obtained over a much large radius. 
This regime therefore has the same focusing properties as in the non-linear regime (for an electron bunch). %
However, the longitudinal wakefields do not attain non-linear values and can therefore be added \emph{linearly} with multiple bunches, possibly allowing for a large transformer ratio or large energy extraction efficiency. %
The addition of the wakefields and the possibility of reaching a large transformer ratio, comparable with that predicted by simple linear PWFA theory has been demonstrated in numerical simulations~\cite{fangR,Rosenzweig10,Rosenzweig12}. %
However, the charge and position of each bunch must be studied using simulations, to produce optimum results. %

\section{Self-modulation instability}

Most PWFA experiments are performed with single electron bunches or short preformed trains of electron bunches. %

Electron bunches can be compressed to extremely short lengths and focused to small transverse sizes, so that PWFA experiments can be performed at high plasma densities ($k_{\mathrm{pe}}\sigma_z\cong\sqrt{2}$, $k_{\mathrm{pe}}\sigma_r<1$) and operate at high gradients. %
However, electron bunches carry relatively small amounts of energy as compared with proton bunches. %
For example a typical SLAC electron bunch with $25\UGeV$ per particle and $2\times10^{10}$ particles per bunch carries about $80\UJ$. %
A typical CERN SPS proton bunch with $400\UGeV$ per particle and $3\times10^{11}$ particles per bunch carries about $19\UkJ$. %
Once accelerated in the LHC to $7\UTeV$, it carries $336\UkJ$. %

A future electron/positron linear collider is expected to produce bunches with at least $250\UGeV$ per particle and $2\times10^{10}$ particles per bunch, \ie carrying $800\UJ$. %
A PBPA-based collider using an $80\UJ$ SLAC-like drive bunch would therefore need to stage at least 10 plasma sections to produce an $800\UJ$ bunch. %
Staging has been envisaged since the beginning of the PWFA. %
However, staging implies distances between the plasma sections that might be much longer than the plasma sections, leading to gradient dilution: the average accelerating gradient (final energy divided by accelerator length or energy gain per plasma stage divided by distance between stages) might be significantly lower than the peak accelerating gradient (gradient in the plasma sections). %

A possible alternative is to use a proton bunch as a drive bunch. %
This idea was explored numerically~\cite{caldwell09}. %
While results showed the possibility of an energy gain as high as ${\sim}600\UGeV$ in a single ${\sim}500\Um$ plasma, the $1\UTeV$, $\sigma_z=100\Uum$, proton bunch used as a driver does not exist. %
CERN SPS and LHC bunches are ${\sim}12\Ucm$ long. %
It is interesting to note that while with electrons or more generally negatively charged bunches, a bunch with $\sigma_r\le\sigma_z\le c/\omega_{\mathrm{pe}}$ is optimum, a shape with $\sigma_z<\sigma_r$ seems more effective for positively charged bunches. %

It was recently proposed to use a transverse instability, the self-modulation instability (SMI) to drive large amplitude wakefields with bunches much longer than the plasma wavelength~\cite{kumar}. %
The SMI arises from the interplay between the transverse wakefields that can periodically focus (or defocus) the particle bunches and increase (or decrease) the bunch density that drives the wakefields more strongly (or weakly) (see Eq.~(\ref{eq:MultiBunches_tran1})). %
The instability modulates the long bunch into a train of bunches with a period approximately equal to the plasma wavelength. %
The train then resonantly drives wakefields to much larger amplitudes than the long bunch would in a plasma with lower density, such that $k_{\mathrm{pe}}\sigma_z\cong\sqrt{2}$. %
One can estimate the \emph{gain} in wakefield amplitude when using SMI by taking $E_{\mathrm{WB}}$ as the measure of the maximum possible accelerating field and assuming that the SMI produces micro-bunches of length such that $k_{\mathrm{pe}}\sigma_r\cong\sqrt{2}$ separated by $\lambda_{\mathrm{pe}}$. %

It is important to note that the modulation has a longitudinal period as a result of the transverse (and not longitudinal) motion of the bunch particles (since dephasing caused by energy gain or loss is negligible for relativistic particles). %

Experiments with electron bunches of one to five  $\lambda_{\mathrm{pe}}$ showed that periodic wakefields are indeed driven by bunches with $\lambda_{\mathrm{pe}}>\sigma_z$~\cite{fang14}. %
The observation was based on longitudinal wakefields; however, those are always accompanied by transverse wakefields. %

Experiments aimed at directly observing the effect of the transverse wakefields on the bunch radius are ongoing at SLAC with electron and positron bunches\cite{vieira14}. %
Preliminary results show evidence of radial modulation, formation of a beam halo and energy loss to wakefields. %
Experiments with low-energy electron bunches are planned at DESY Zeuthen~\cite{gross14}. %

A major experiment, known as AWAKE, is in the design and installation phase at CERN~\cite{AWAKE}. %
It will use the $400\UGeV$ SPS proton bunch and a $10\Um$ long plasma with a baseline density $n_{\mathrm{e}0}=7\times10^{14}\Ucm^{-3}$~\cite{oz14}. %

The first goal of the experiment is to observe and characterize the SMI of the proton bunch. %
In a second phase, an electron bunch with $\sigma_z$ of the order of one to two $\lambda_{\mathrm{pe}}$ will be externally injected to sample the wakefields. %
In a third phase, an electron bunch shorter than $\lambda_{\mathrm{pe}}$, possibly produced by a LWFA~\cite{muggli14}, will be injected to narrow the final energy spread through loading of the wakefields and to preserve the accelerated bunch emittance~\cite{veronika}. %
Details of the AWAKE experiment can be found in the lecture by E. Gschwendtner. %

\section{Final remarks}

The PWFA is one of the advanced accelerator schemes studied as a high-gradient alternative to today's RF technology. %
The PWFA has its advantages and its limits, and only further research in the field will help determine its relevance to a current or future electron or positron accelerator. %
The application of the PWFA is to high-energy accelerators, such as those driving X-ray free electron lasers (FELs) or the electron/positron collider that may be chosen to complement the LHC. %

It is clear that many subjects have not been addressed in this manuscript: plasma sources, staging of plasma sources, bunch head erosion, scattering on the plasma ions, \etc %
However, we hope that this manuscript will foster further reading about and understanding of the growing body of theoretical, numerical and experimental PWFA work. %
Moreover, we hope that curiosity, talent and enthusiasm of many scientists, potentially captured by this CERN accelerator school, will take this concept to its limit. %

\end{document}